\documentclass{ws-ijbc}
\usepackage{graphicx}
\usepackage{enumerate}
\usepackage{color}
\usepackage{url}

\begin{document}

\title{RECURRENCE-BASED TIME SERIES ANALYSIS BY MEANS OF COMPLEX NETWORK METHODS}

\author{REIK V. DONNER$^{1,2,3}$\thanks{Electronic address: \texttt{reik.donner@pik-potsdam.de}}, 
MICHAEL SMALL$^4$\thanks{Electronic address: \texttt{ensmall@polyu.edu.hk}}, 
JONATHAN F. DONGES$^{2,5}$\thanks{Electronic address: \texttt{donges@pik-potsdam.de}}, 
NORBERT MARWAN$^2$\thanks{Electronic address: \texttt{marwan@pik-potsdam.de}}, 
YONG ZOU$^2$\thanks{Electronic address: \texttt{yong.zou@pik-potsdam.de}}, 
RUOXI XIANG$^4$\thanks{Electronic address: \texttt{ms.ruoxi-xiang@polyu.edu.hk}}, and 
J\"URGEN KURTHS$^{2,5}$\thanks{Electronic address: \texttt{juergen.kurths@pik-potsdam.de}}}

\address{$^1$Max Planck Institute for Physics of Complex Systems, N\"othnitzer Str.~38, 01187 Dresden, Germany \\ $^2$Potsdam Institute for Climate Impact Research, P.O. Box 60\,12\,03, 14412 Potsdam, Germany \\
$^3$Institute for Transport and Economics, Dresden University of Technology, W\"urzburger Str.~35, 01187 Dresden, Germany \\
$^4$Department of Electronic and Information Engineering, Hong Kong Polytechnic University, Hung Hom, Kowloon, Hong Kong \\
$^5$Department of Physics, Humboldt University Berlin, Newtonstr.~15, 12489 Berlin, Germany}

\date{\today}

\maketitle

\begin{abstract}

Complex networks are an important paradigm of modern complex systems sciences which allows quantitatively assessing the structural properties of systems composed of different interacting entities. During the last years, intensive efforts have been spent on applying network-based concepts also for the analysis of dynamically relevant higher-order statistical properties of time series. Notably, many corresponding approaches are closely related with the concept of recurrence in phase space. In this paper, we review recent methodological advances in time series analysis based on complex networks, with a special emphasis on methods founded on recurrence plots. The potentials and limitations of the individual methods are discussed and illustrated for paradigmatic examples of dynamical systems as well as for real-world time series. Complex network measures are shown to provide information about structural features of dynamical systems that {are complementary to those characterized by other methods of time series analysis and, hence,} substantially enrich the knowledge gathered from other existing (linear as well as nonlinear) {approaches}.


\end{abstract}



\section{Introduction} \label{sec:introduction}

The understanding of principles and mechanisms underlying the dynamics of natural systems is closely related to the progress of complex systems analysis. Concepts originated in the field of nonlinear dynamics such as correlation dimension~\cite{grassberger83} or Lyapunov exponents~\cite{wolf85} have been introduced and successfully applied for quantitatively describing phase space topology and resulting dynamical properties. Spatially extended systems have been studied using, {\textit{e.g.,}} fractal properties~\cite{marwan2007epjst,Dombradi2007}, information-based measures~\cite{schirdewan2007}, or complex network approaches~\cite{Donges2009b}.

In the past two decades, a new class of dynamical characteristics has received increasing attention, which is based on the widely observed phenomenon of recurrences~\cite{marwan2007}. Many dynamical processes exhibit recurrences, which have already been recognized by \citet{poincare1890} in his seminal study of the three-body problem. In the context of time series analysis, we refer to a recurrence of a state $\vec{x}_i$ at time $t = i\cdot \Delta t$ (where $i \in \mathbb{N}$, $\Delta t $ is the sampling time, and $\vec{x} \in \mathbb{R}^m$ a state in the $m$-dimensional phase-space) whenever the state of the system $\vec{x}_j$ at another time $j\cdot \Delta t$ is similar to that initial state (\textit{i.e.,} $\vec{x}_i \approx \vec{x}_j$) or as close as we wish (but usually not identical)\footnote{Whenever we refer to a state in phase-space, we consider either a system with all known system variables, or a phase space which is reconstructed from a time series, \textit{e.g.,} by means of time-delay embedding~\cite{packard80,Takens1981}. In the following, we presume that the reader is familiar with embedding techniques and estimation of embedding parameters.}. Despite the implicit technical restriction of constant sampling time made in this definition, we would like to note that unlike some other basic approaches of time series analysis, the recurrence concept can be directly generalized to unequally sampled data. 

The increasing interest in using the concept of recurrence for the analysis of dynamical systems is related to the introduction of more and more powerful computers~\cite{marwan2008epjst}. First return maps and recurrence time statistics have been introduced to study chaotic dynamical systems, unstable periodic orbits, or dynamical invariants~\cite{procaccia1987,gao99}. \citet{Eckmann1987} have introduced recurrence plots (RPs) for visualization of recurrences in phase space. A RP represents all recurrences in form of a binary matrix $\mathbf{R}$, where $R_{i,j}=1$ if the state $\vec{x}_j$ is a neighbor of $\vec{x}_i$ in phase space, and $R_{i,j}=0$ otherwise.

\begin{figure}[thb]
\centering
\includegraphics[width=\textwidth]{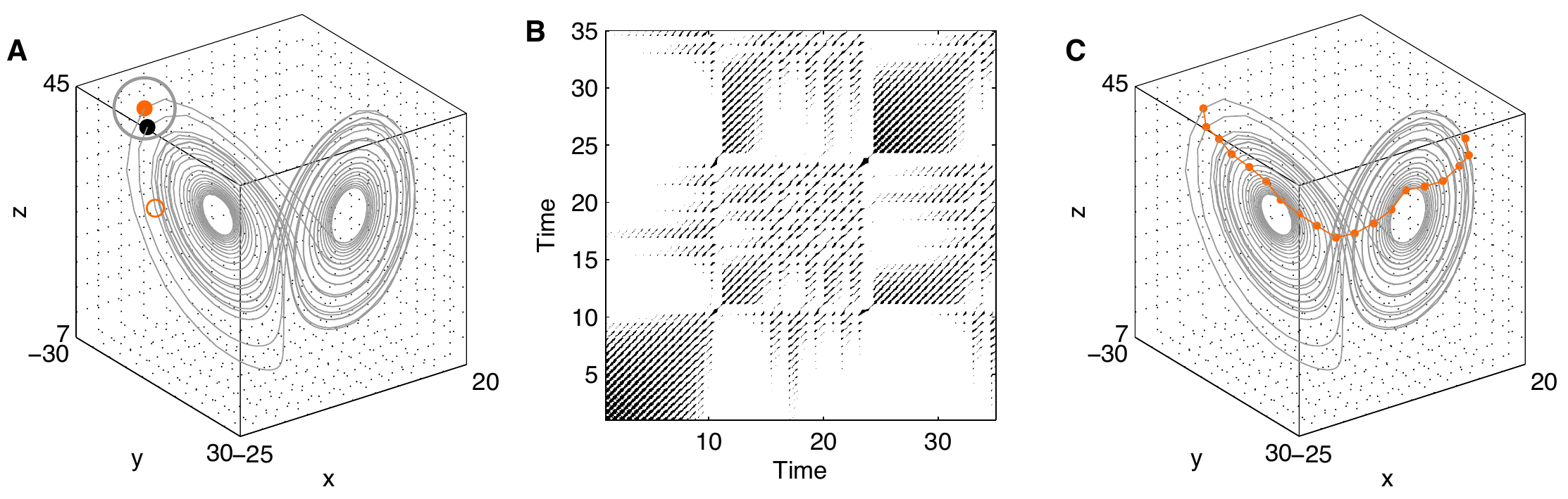}
\caption{Basic concepts beyond recurrence plots and the resulting recurrence networks, exemplified for one realization of the Lorenz system (Eq.~(\ref{eq_lorenz})) with the parameters $r=28$, $\sigma=10$ and $\beta=8/3$ (sampling time $\Delta t=0.02$, original coordinates, no embedding, recurrences defined based on a fixed threshold $\varepsilon=5.0$ using maximum norm). (A) A state at time $i$ (red dot) is recurrent at another time $j$ (black dot) when the phase space trajectory visits its close neighborhood (gray circle). This is marked by value 1 in the recurrence matrix at $(i,j)$. States outside of this neighborhood (small red circle) are marked with 0 in the recurrence matrix. (B) Graphical representation of the corresponding recurrence matrix (recurrence plot) and adjacency matrix (modulo main diagonal). (C) A particular path in the recurrence network for the same system embedded in the corresponding phase space.}
\label{lorenz_constr}
\end{figure}

RPs can be defined in different ways. In the original RP definition of \citet{Eckmann1987}, only the $k$ nearest neighbors of states in phase space are considered. This preserves a constant column sum in $\mathbf{R}$, \textit{i.e.,} the recurrence point density (or local recurrence rate)
\begin{equation}
RR_i=\frac{1}{N}\sum_{j=1}^N R_{i,j}
\end{equation}
\noindent
is conserved at $RR_i=k/N$ (with $N$ being the length of the time series). The advantage of this method is that it allows comparing RPs of different systems without the necessity of normalizing the underlying time series beforehand, since the global recurrence rate
\begin{equation}
RR=\frac{1}{N^2}\sum_{i,j=1}^N R_{i,j}
\end{equation}
\noindent
is fixed at the same value. Alternatively, in the most common definition of a RP, a state is considered to be recurrent if the system's trajectory approaches state $\vec{x}_i$ in phase space closer than a certain recurrence threshold $\varepsilon$, \textit{i.e.,}
\begin{equation}\label{eq_rp}
R_{i,j}(\varepsilon) = 
\Theta\left(\varepsilon-\left\|\vec x_{i} - \vec x_{j}\right\|\right),
\end{equation}
where $\Theta (\cdot)$ is the Heaviside function and $\left\| \cdot \right\|$ is a norm. The basic principle is illustrated in Fig.~\ref{lorenz_constr} for one realization of the Lorenz system 
\begin{equation}
\frac{d}{dt}\left(\begin{array}{c}x\\y\\z\end{array}\right)=\left(\begin{array}{c}\sigma(y-x) \\ x(r-y)\\ xy-\beta z \end{array}\right).
\label{eq_lorenz}
\end{equation}
\noindent
Further definitions of recurrences add dynamical aspects, such as local rank orders or strictly parallel evolution of states (parallel segments of phase-space trajectory considered in iso-directional RPs~\cite{Horai2002}). For a more detailed overview, we refer to~\cite{marwan2007,bandt2008}.

\begin{figure}[thb]
\centering
\includegraphics[width=\textwidth]{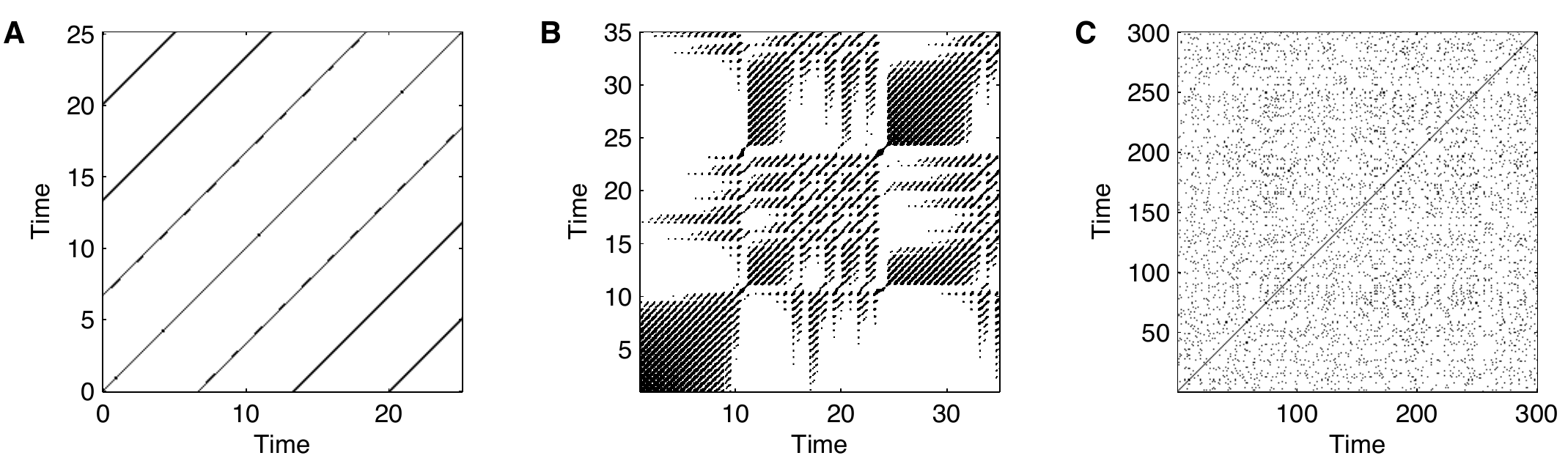}
\caption{Exemplary recurrence plots of (A) a periodic motion with one frequency, (B) the chaotic Lorenz system (same trajectory as in Fig.~\ref{lorenz_constr}A), and (C) of normally distributed white noise.}\label{rp_examples}
\end{figure}

RPs of dynamical systems with different types of dynamics exhibit distinct structural properties (see Fig.~\ref{rp_examples}), which can be characterized in terms of their associated small-scale as well as large-scale features~\cite{marwan2007}. A periodic regime is reflected by long and non-interrupted diagonal lines. The vertical distance between these lines corresponds to the period of the oscillation. A chaotic dynamics also leads to diagonals, which are however clearly shorter. There are also certain vertical structures, which are not as regular as in the case of a periodic motion. The RP of an uncorrelated stochastic signal consists of mainly isolated black points. The distribution of the points in such a RP appears rather erratic but nevertheless homogeneous.

The study of recurrences by means of RPs has become popular with the introduction of recurrence quantification analysis (RQA)~\cite{zbilut92,marwan2002herz}. The initial purpose of this framework has been to introduce measures of complexity which distinguish between different appearances of RPs~\cite{marwan2008epjst}, since they are linked to certain dynamical properties of the studied system. RQA measures use the distribution of small-scale features in the RP, namely individual recurrence points as well as diagonal and vertical line structures. RQA as a whole has been proven to constitute a very powerful technique for quantifying differences in the dynamics of complex systems and has meanwhile found numerous applications, \textit{e.g.,} in astrophysics~\cite{kurths94}, ecology~\cite{facchini2007}, engineering~\cite{litak2009c}, geo- and life sciences~\cite{marwan2003climdyn,marwan2007pla}, or protein research~\cite{Giuliani2002a,zbilut2004a}. For a more comprehensive
review on the potentials of this method, we refer to~\cite{marwan2008epjst,webber2009}. In addition, we would like to remark that even dynamical invariants, like the $K_2$ entropy and mutual information, or dimensions (information and correlation dimensions $D_1$, $D_2$) can be efficiently estimated from RPs~\cite{thiel2004a,marwan2007}. Moreover, RPs have also been successfully applied to study interrelations, couplings, and phase synchronization between dynamical systems~\cite{marwan2002npg,romano2004,romano2005,romano2007,vanLeeuwen2009,nawrath2010}.

Another appealing concept for analyzing structural features of complex systems is based on their representation as complex networks of passive or active (\textit{i.e.,} mutually interacting) subsystems. {An undirected, unweighted} complex network $G$, consisting of $N$ vertices and $E$ edges, is conveniently represented by the binary adjacency matrix $\mathbf{A}$, where $A_{i,j}=1$ if vertex $i$ connects to vertex $j$, and $A_{i,j}=0$ if the edge $(i,j)$ does not exist.

Starting from mathematical results on graph theory, numerous applications of complex networks have been considered in the literature, including studies of networked infrastructures \cite{Amaral2000,Latora2001,Guimera2005}, the derivation of network patterns from empirical data of social interactions \cite{Freeman1979}, the assessment of functional connectivity in the brain from spatially distributed (multi-channel) neurophysiological measurements \cite{Zhou2006,Zhou2007}, or the identification of dynamically relevant backbone structures in complex network representations of continuous systems such as atmospheric dynamics \cite{Donges2009a,Donges2009b}, to mention only some important recent fields of application. For a more detailed statistical description of the topological features of real-world as well as model networks, a large variety of different statistical measures have been suggested~\cite{Albert2002,Newman2003,daCosta2007}. These measures have been successfully applied to quantify the properties of complex networks in various scientific disciplines, fostering substantial progress in our understanding of the interplay between structure and dynamics of such networks \cite{Wang2002,Boccaletti2006,Arenas2008}.

\begin{figure}[thb]
\centering
\includegraphics[width=0.48\textwidth]{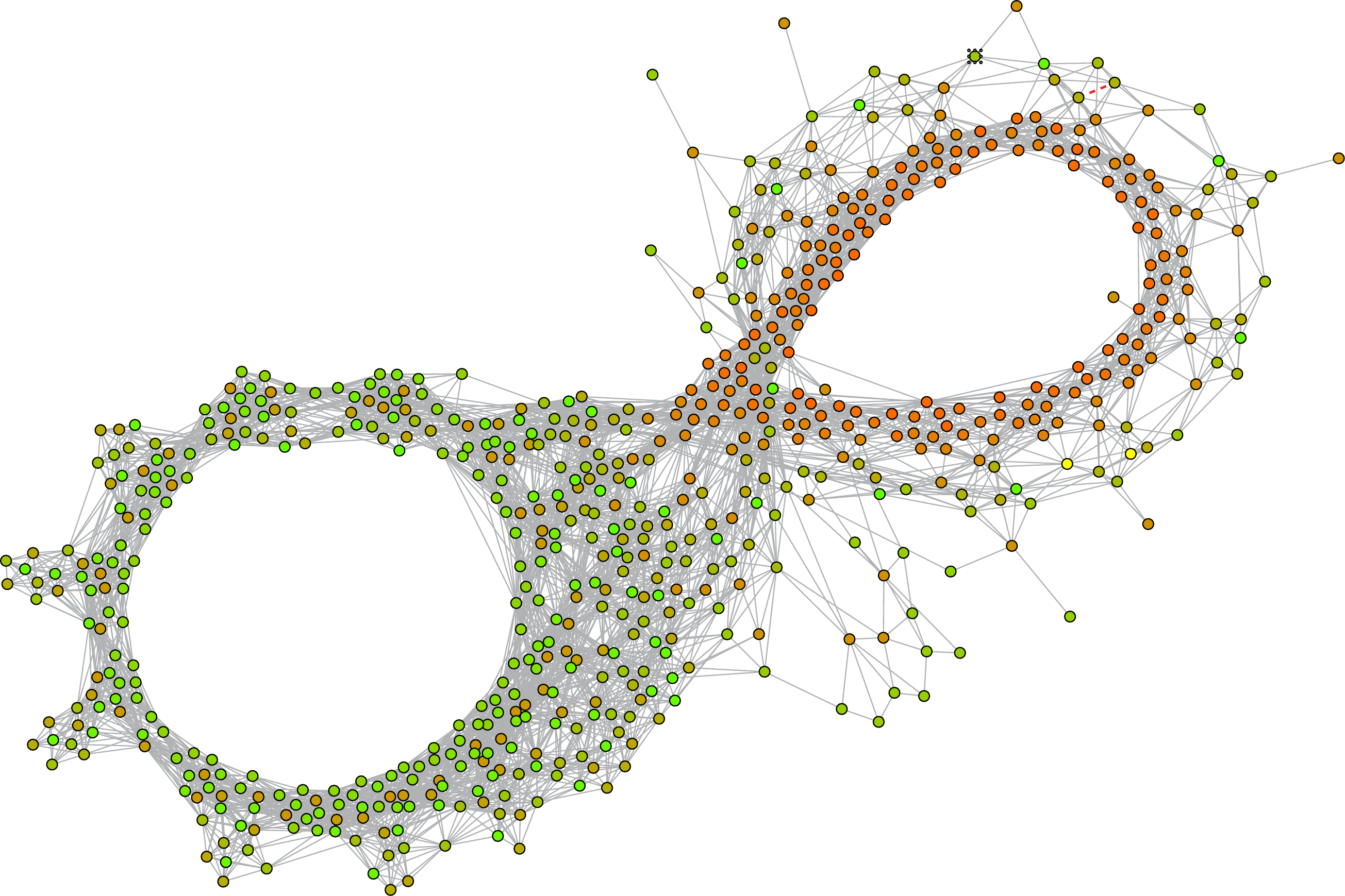}
\caption{A graphical representation of the Lorenz attractor based on the recurrence matrix represented in Fig.~\ref{lorenz_constr}. The color of the vertices corresponds to their temporal order (from orange to bright green).}
\label{lorenz_net}
\end{figure}

We emphasize that there are strong conceptual similarities between, on the one hand, the reconstruction of network topologies from spatially distributed time series (\textit{e.g.,} in neurophysiological or climate networks) and, on the other hand, the study of phase space properties of dynamical systems based on individual time series. Following this idea, fundamental characteristics of a dynamical system can be captured by properly defining complex networks based on such time series. Among other methods, the re-interpretation of the recurrence matrix $\mathbf{R}$ as the adjacency matrix $\mathbf{A}$ of an unweighted complex network (Figs.~\ref{lorenz_constr} and \ref{lorenz_net}) provides a novel concept for nonlinear time series analysis~\cite{Marwan2009,Donner2009,Donner2010}.

The remaining part of this paper is organized as follows: In Sec.~\ref{sec:tsnet}, we review recent approaches for transforming time series into complex network representations. Specifically, complex networks based on different definitions of RPs (so-called recurrence networks) are discussed in some detail. Section~\ref{sec:technical} summarizes technical issues that have to be considered when systematically applying the different approaches to time series analysis. Finally, two examples for real-world applications of recurrence networks are discussed in Sec.~\ref{sec:applications}.

\section{Transforming time series into complex networks} \label{sec:tsnet}

Recently, several approaches have been proposed for transforming (observational) time series into complex network representations. These methods can be roughly distinguished into three classes (see Tab.~\ref{tab:methods}), which are based on
\begin{enumerate}[(i)]
\item mutual proximity of different segments of a time series (proximity networks),
\item convexity of successive observations (visibility graphs), and
\item transition probabilities between discrete states (transition networks).
\end{enumerate}
\noindent
With the exception of visibility graphs, all approaches are related with the concept of recurrence. This is particularly evident for proximity networks, where connectivity is defined in a data-adaptive local way, \textit{i.e.,} by considering distinct regions with a varying center at a given vertex in either the phase space itself or an abstract proximity space. In contrast, for transition networks, the corresponding classes are rigid, \textit{i.e.,} determined by a fixed coarse-graining of the phase space. The distinction between both classes of approaches is conceptually similar to the duality of symbolic time series analysis (\textit{i.e.,} time series analysis based on a coarse-graining of the dynamics) and quantitative analysis of RPs~\cite{Donner2008}, which may both be used for estimating similar dynamical invariants such as entropies and mutual information.

\begin{table}[t]
\tbl{Summary of the definitions of vertices and the criteria for the existence of edges in existing complex network approaches to time series analysis.}{
\small
\begin{tabular}{llll}
\hline
Method & Vertex & Edge & Directedness \\
\hline
Proximity networks & & \\
\textit{Cycle networks} & Cycle & Correlation or phase space distance between cycles & undirected \\
\textit{Correlation networks} & State vector & Correlation coefficient between state vectors & undirected \\
\textit{Recurrence networks} & & & \\
\quad \textit{$k$-nearest neighbor networks}& State (vector) & Recurrence of states (fixed neighborhood mass) & directed \\
\quad \textit{adaptive nearest neighbor networks}& State (vector)  & Recurrence of states (fixed number of edges) & undirected \\
\quad \textit{$\varepsilon$-recurrence networks} & State (vector) & Recurrence of states (fixed neighborhood volume) & undirected \\
\hline
Visibility graphs & Scalar state & Mutual visibility of states & undirected \\
\hline
Transition networks & Discrete state & Transitions between states & directed \\
\hline
\end{tabular}
\normalsize
\label{tab:methods}}
\end{table}

In addition to these specific relationships between the recurrence concept and different types of time series networks, there is a fundamental structural analogy between RPs and (unweighted) complex networks in general. Both structures are based on binary matrices (\textit{i.e.,} recurrence and adjacency matrices, respectively) that can be used for studying basic topological properties of the underlying complex system based on sophisticated statistical measures. Proximity and transition networks as well as RPs based on Eq.~(\ref{eq_rp}) can be generalized by withdrawing the application of a specific threshold, which leads to weighted networks and unthresholded RPs (distance plots), respectively. For example, the unthresholded RP obtained from one trajectory of a given dynamical system may be re-interpreted as the connectivity matrix of a fully coupled, weighted network.

Among the three classes of methods listed above, the largest group of concepts is given by proximity networks, where the mutual closeness or similarity of different segments of a trajectory can be characterized in different ways. Consequently, there are different types of such proximity networks (see Tab.~\ref{tab:methods}): cycle networks, correlation networks, and recurrence networks. However, all these methods are characterized by two common general properties:

Firstly, the resulting networks are invariant under relabeling of their vertices in the adjacency matrix. Hence, the topological characteristics of proximity networks yield nonlinear measures that are invariant against permutation of vertices. In this respect, the network-theoretic approach is distinctively different from traditional methods of time series analysis where the temporal order of observations does explicitly matter. 

Secondly, we have to point out that particularly proximity networks are spatial networks. In particular, recurrence networks are embedded in the phase space of the considered system, with distances being defined by one of the standard metrics (\textit{e.g.,} Euclidean, Manhattan, etc.). Similar considerations apply to other types of proximity networks as well. 

Both mentioned characteristics imply that the network-theoretic concept of a path on a given graph (see Fig.~\ref{lorenz_constr}C) is distinctively different from the trajectory concept that records the causal dynamic evolution of the system~\cite{Donner2009}. Note that unlike for proximity networks, causal relationships are conserved in transition networks (and at least to some extent also in visibility graphs).

In the following, we will discuss the main properties of the different concepts in some detail.

\subsection{Cycle networks} 

\citet{Zhang2006} (see also \cite{Zhang2008,Small2009}) {first} suggested to study the topological features of pseudo-periodic time series by means of complex networks. Suppose that a dynamical system possesses pronounced oscillations (examples are the well-known Lorenz and R\"ossler systems). In this case, we identify the individual cycles contained in a time series of this system with the vertices of an undirected network. Edges between pairs of vertices are established if the corresponding segments of the trajectory behave very similarly. For quantifying the proximity of cycles in phase space, different measures have been proposed. \citet{Zhang2006PRE} introduced a generalization of the correlation coefficient applicable to cycles of possibly different lengths. Specifically, this correlation index is defined as the maximum of the cross correlation between the two signals when the shorter of both is slid relative to the longer one. That is, if the two cycles being compared are $C_1=\{x_1,x_2,\ldots,x_{\alpha}\}$ and $C_2=\{y_1,y_2,\ldots,y_{\beta}\}$ with (without loss of generality) $\alpha\leq\beta$, then we compute 
\begin{equation}
\rho(C_1,C_2)=\max_{i=0,\ldots (\beta-\alpha)} \left<(x_1,x_2,\ldots,x_{\alpha}),(y_{1+i},y_{2+i},\ldots, y_{\alpha+i})\right>,
\label{cyclecorr}
\end{equation}
\noindent
where $\left<\cdot,\cdot\right>$ denotes the standard correlation coefficient of two $\alpha$-dimensional vectors, and set
\begin{equation}
A_{i,j}=\Theta(\rho(C_i,C_j)-\rho_{max})-\delta_{i,j}.
\end{equation}
\noindent 
where $\delta_{i,j}$ is the Kronecker delta necessary in order to obtain a network without self-loops. As an alternative, the phase space distance~\cite{Zhang2006}
\begin{equation}
D(C_1,C_2)=\min_{i=0,\ldots (\beta-\alpha)} \frac{1}{\alpha} \sum_{j=1}^{\alpha} \|x_j-y_{j+i}\|
\label{psd}
\end{equation}
\noindent
has been suggested, leading to the following definition: 
\begin{equation}
A_{i,j}=\Theta(D_{max}-D(C_i,C_j))-\delta_{i,j}.
\end{equation}
\noindent 
Of course, there are other calculations one could perform as well. 

\begin{figure}[thb]
\centering
\includegraphics[width=0.75\textwidth]{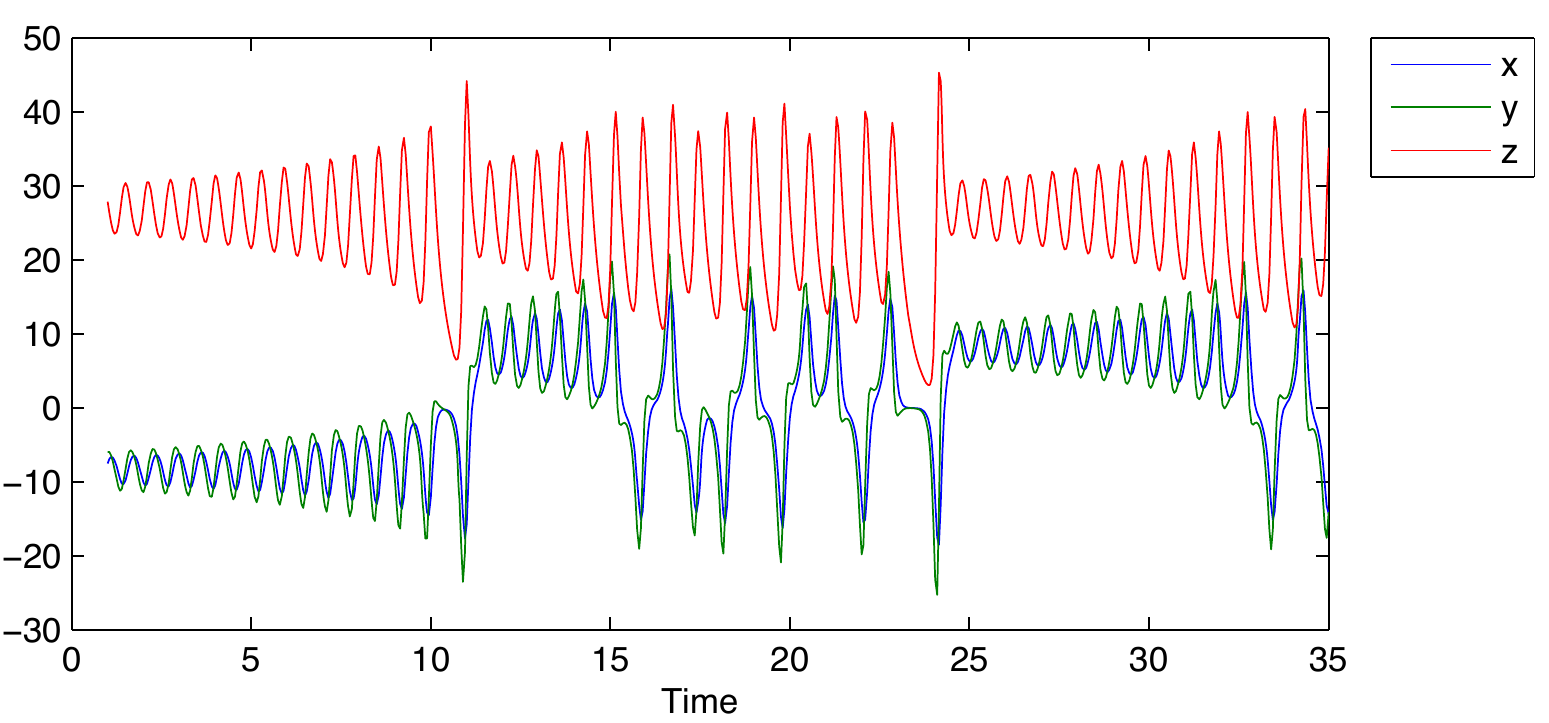}
\caption{Time series of the trajectory of the Lorenz system (Eq.~(\ref{eq_lorenz})) used in Figs.~\ref{lorenz_constr}, \ref{rp_examples}  and \ref{lorenz_net} ($\Delta t=0.05$).}
\label{lorenz_data}
\end{figure}

\begin{figure*}[thb]
\centering
\includegraphics[width=0.6\textwidth]{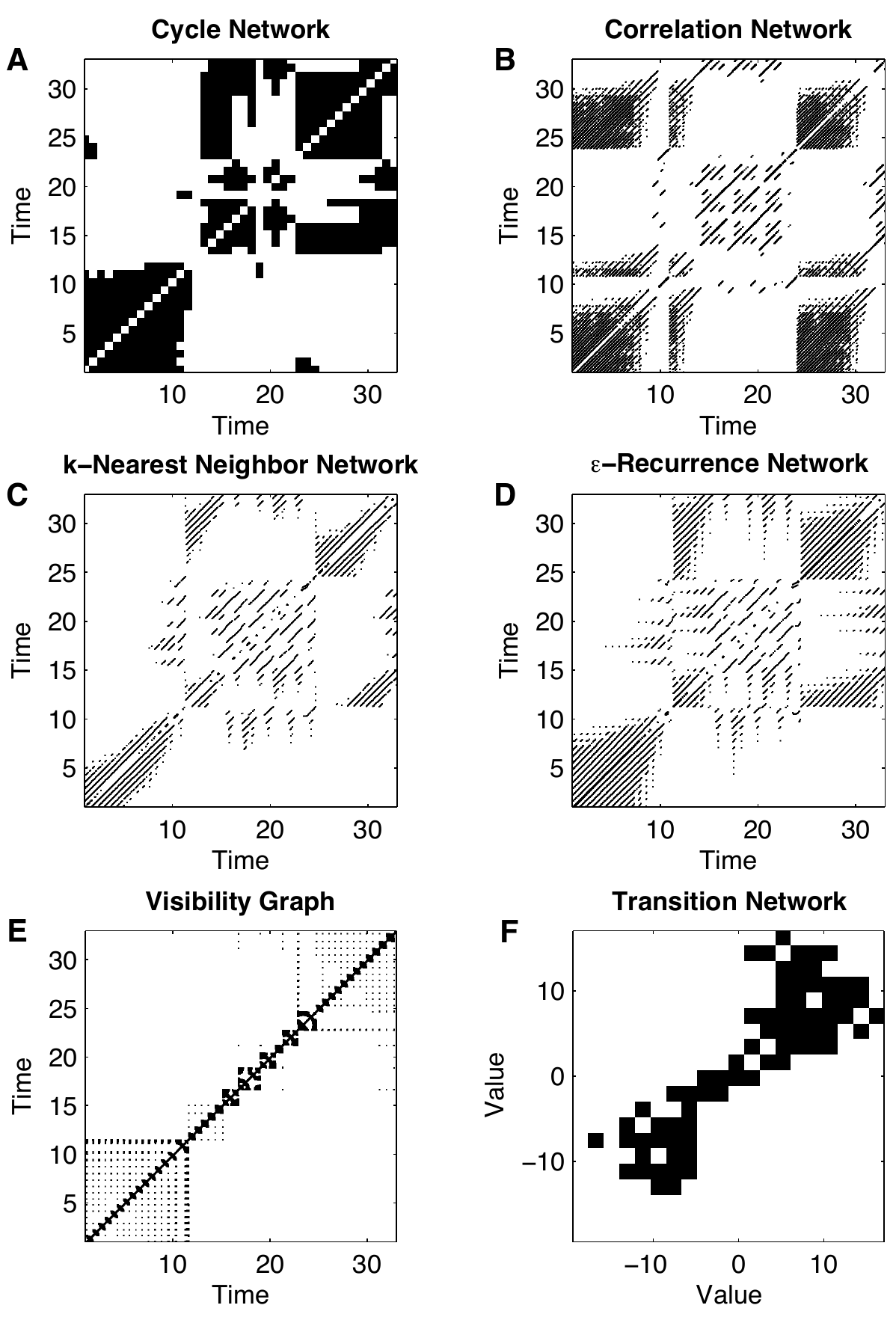}
\caption{Adjacency matrices corresponding to different types of networks constructed from the $x$-coordinate of the Lorenz system shown in Fig.~\ref{lorenz_data}: (A) Cycle network ($N=40$, critical cycle distance in phase space $D_{max}=5$), (B) correlation network ($N=654$, embedding dimension $m=10$ with delay $\tau=3$ time steps), (C) $k$-nearest neighbor network {(asymmetric version)}, $N=675$, $m=3$, $\tau=3$, $k=10$, corresponding to a recurrence rate of $RR\approx 0.015$ using Euclidean norm; the associated adaptive nearest neighbor network (not shown) is characterized by a very similar pattern), (D) $\varepsilon$-recurrence network ($N=675$, $m=3$, $\tau=3$, $\varepsilon=2$, maximum norm), (E) visibility graph ($N=681$), and (F) transition network (based on an equipartition of the range of observed values into $N=20$ classes of size $\Delta x=3.0$, minimum transition probability $p=0.2$ during 3 time steps). Note that {only in panels (C) and (D), the adjacency matrices correspond to recurrence matrices of the underlying time series according to the standard definition~\cite{Eckmann1987,marwan2007}. In both cases}, recurrence points originated from strong tangential motion (sojourn points) have been removed, resulting in additional asymmetries.}
\label{lorenz_linkmatrix}
\end{figure*}

\begin{figure*}[thb]
\centering
\includegraphics[width=0.95\textwidth]{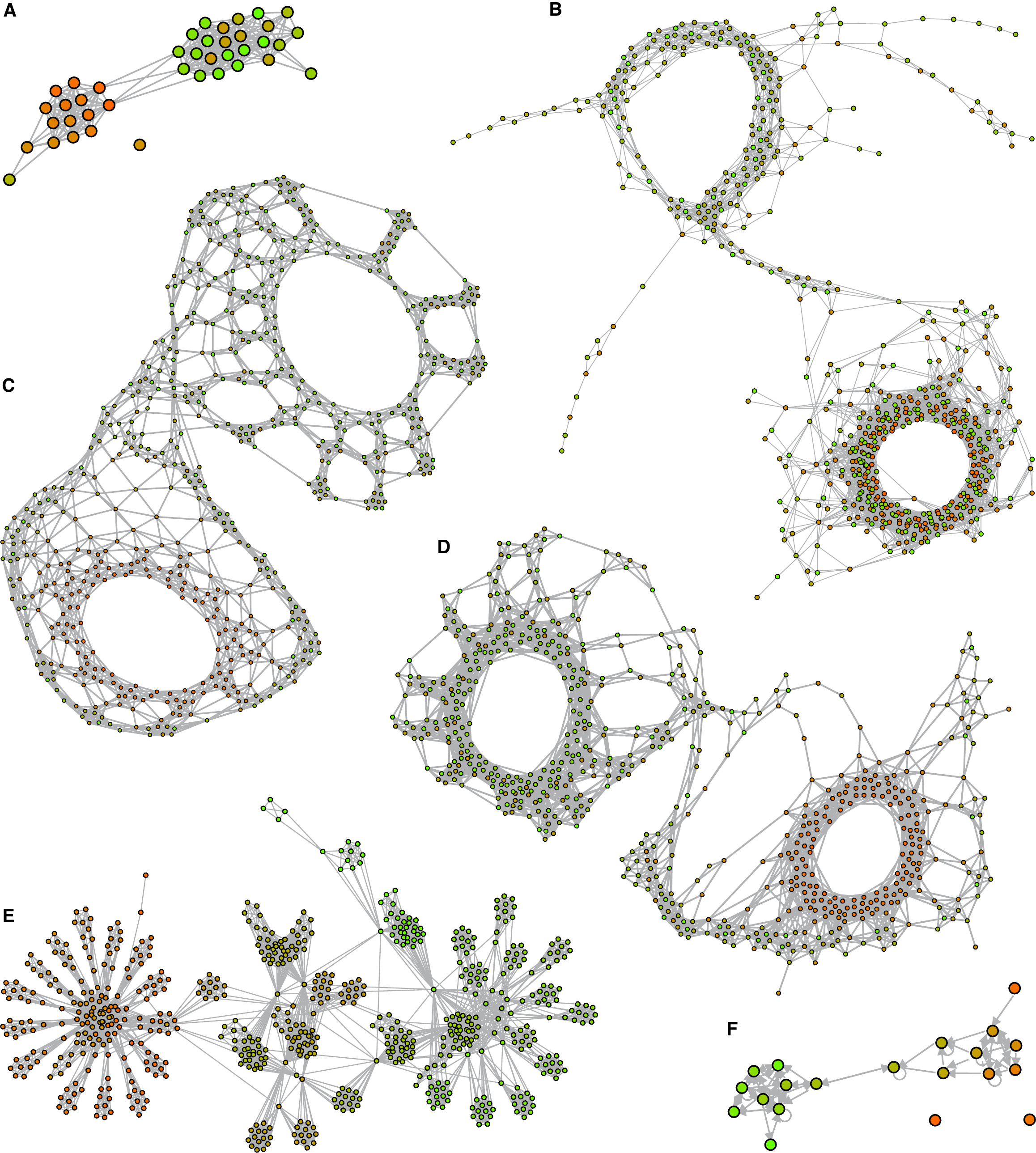}
\caption{Graphical representation of the different complex networks based on the adjacency matrices shown in Fig.~\ref{lorenz_linkmatrix}. The graphs have been embedded into an abstract two-dimensional space using a force directed placement algorithm~\cite{Battista1994}. For panels (A)-(E), the vertex color indicates the temporal order of observations (from orange to bright green), for the transition network (panel (F)), colors correspond to the different $x$ values. Note that in panels (B) and (D), some individual disconnected vertices have been removed from these network representations.}
\label{lorenz_networks}
\end{figure*}

As an example for constructing complex networks from time series, Fig.~\ref{lorenz_data} shows one realization of the Lorenz system, which is characterized by a double-scroll topology of the attractor with pronounced chaotic oscillations. For the first about 10 time units of simulation, the system rotates around one of both unstable centers ($x,y<0$), is then captured by the other part of the attractor ($x,y>0$) for about 5 more time units, followed by some fast transitions between both parts (\textit{i.e.,} involving only one or two subsequent oscillations around each center), before performing again rotations around the second center ($x,y>0$) between $t\approx 25$ and $t\approx 32$. This structure is well represented in the adjacency matrix $\mathbf{A}$ of the corresponding cycle network based on the $x$-coordinate time series (see Fig.~\ref{lorenz_linkmatrix}A), where we observe several pronounced clusters along the main diagonal corresponding to the two distinct parts of the attractor. Consequently, the resulting network (Fig.~\ref{lorenz_networks}A) shows a pronounced community structure with two groups, corresponding to the double-scroll topology of the system.

The advantage of cycle networks is that {explicit} time delay embedding is avoided. {In addition,} the method is {more robust than other methods} against additive noise, {given a small enough noise magnitude to allow a clear identification of the individual cycles from the time series}. Moreover, {cycle networks are} invariant under reordering of the cycles (this is precisely the same property that was also exploited for cycle-shuffled surrogate methods~\cite{Theiler1996} but not the pseudo-periodic surrogate method~\cite{Small2001PRL}). However, for chaotic and nonlinear systems in a near-periodic regime, we typically observe significant orderly variation in the appearance of individual cycles. For systems that are linear or noise driven, that orderly variation will be less pronounced. As a consequence, the networks constructed with these methods will have characteristic and distinct properties: linear and periodic systems have cycle networks that appear randomly, while chaotic and nonlinear systems generate highly structured networks \cite{Zhang2006,Zhang2008}. Therefore, the vertex and edge properties of the resultant networks can be used to distinguish between distinct classes of dynamical systems. Moreover, \citet{Zhang2006} used meso-scale properties of the networks --- and in particular the clustering of vertices --- to locate unstable periodic orbits (UPOs) within the system. This approach is feasible, since a chaotic system will exhibit a dense hierarchy of unstable periodic orbits, and these orbits act as accumulation points in the Poincar\'e section. Hence, the corresponding vertices form clusters in the cycle network.

\subsection{Correlation networks} 

By embedding an arbitrary time series, individual state vectors $\vec{x}_i$ in the $m$-dimensional phase space of the embedded variables can be considered as vertices of an undirected complex network. Specifically, if the Pearson correlation coefficient 
\begin{equation}
r_{i,j} = \left<\vec{x}_i,\vec{x}_j\right>
\end{equation}
\noindent
is larger than a given threshold $r$, the vertices $i$ and $j$ are considered to be connected \cite{Yang2008,Gao2009}:
\begin{equation}
A_{i,j}=\Theta(r-r_{i,j})-\delta_{i,j}.
\end{equation}
\noindent 
Interpreting $1-r_{i,j}$ as a proximity measure, the condition $r_{i,j}\geq r$ corresponds to the definition (\ref{eq_rp}) of a recurrence with $\varepsilon=1-r$. The consideration of correlation coefficients between two phase space vectors usually requires a sufficiently large embedding dimension $m$ for a proper estimation of $r_{i,j}$. Hence, information about the short-term dynamics might get lost. Moreover, since embedding is known to induce spurious correlations~{\cite{Thiel2006b}}, the results of the correlation method of network construction may suffer from related effects.

The adjacency matrix of a correlation network is shown in Fig.~\ref{lorenz_linkmatrix}B for the same example trajectory of the Lorenz system as previously used for constructing a cycle network. In contrast to the cycle network, we observe strong connectivity between vertices corresponding to the time intervals $t=0\dots 7$ and $t=24\dots 28$, \textit{i.e.,} two time intervals where the trajectory is actually captured in two different parts of the attractor. An explanation for this behavior is that the dynamics itself within the two time intervals appears to be rather similar (see Fig.~\ref{lorenz_data}), but it is just shifted in the $x$- (and $y$-) coordinate. Since the estimation of correlation coefficients between embedding vectors explicitly removes the mean position of the trajectory during the different time intervals covered by these vectors, the two respective parts of the trajectory are considered to be similar with respect to the correlation criterion, although they are actually well separated in the actual phase space. This example underlines that correlations must be carefully distinguished from true metric distances.

Visualization of the correlation network embedded in an abstract two-dimensional space (Fig.~\ref{lorenz_networks}B) reveals a pronounced community structure with two major groups that are characterized by a ring-like topology. However, these two groups do not correspond to the two scrolls of the attractor, as is the case for the cycle network and also for recurrence networks (see Sec.~\ref{sec:recnet_intro}). In contrast, it is a reasonable assumption that the observed group structure is determined by the orientation of the arc-like embedding vectors around each of the centers.

\subsection{Recurrence networks} \label{sec:recnet_intro}

A recurrence network is a complex network whose adjacency matrix is given by the recurrence matrix of a time series\footnote{In \cite{Donner2009,Donner2010}, the term recurrence network has been more specifically used for the special case in which recurrences have been defined according to (\ref{eq_rp}).}, \textit{i.e.,} we define the adjacency matrix of a recurrence network by
\begin{equation}
A_{i,j}=R_{i,j}-\delta_{i,j}.
\label{recnetdef}
\end{equation}
\noindent
Note that removing the line of identity from the RP corresponds to the consideration of the smallest possible Theiler window in traditional RQA~\cite{marwan2007}. 

Since information about the temporal ordering of observations is not explicitly regarded in a recurrence network defined according to {Eq.}~(\ref{recnetdef}), the topological features of the resulting graphs reflect dynamically invariant properties associated with the specific dynamical system. From this perspective, the quantitative analysis of recurrence networks, although being based on the same recurrence matrix as traditional RQA, reflects distinctively different properties of the system than line-based RQA measures. Hence, besides RQA and the estimation of dynamical invariants based on line structures in RPs, the analysis of recurrence networks can be considered as a third column for the quantitative recurrence-based characterization of phase space properties of dynamical systems. Moreover, while the appropriate estimation of most RQA measures requires the careful choice of a second parameter (the minimum line length $l_{min}$), quantitative characteristics of recurrence networks involve only a single parameter (depending on the specific algorithm, see below). However, computing network-theoretic measures (\textit{e.g.,} betweenness centrality)~\cite{Newman2003} often requires larger computational efforts than traditional RQA.

Since the recurrence matrix can be defined in different ways (see Sec.~\ref{sec:introduction}), there are distinct sub-types of recurrence networks that are characterized by somewhat different structural properties:

\subsubsection{$k$-nearest neighbor networks}

Following the original definition of a RP by \citet{Eckmann1987}, every (possibly embedded) observation vector is considered as a vertex $i$, which is then linked to those $k$ other vertices $j$ that have the shortest mutual distances $d_{i,j}$ with respect to $i$ in phase space (\textit{i.e.,} to its $k$ nearest neighbors). This means that a \textit{directed} edge is introduced from $i$ to every vertex $j\in\mathcal{N}^{(k)}_i$, where $\mathcal{N}^{(k)}_i$ is the set of $k$ nearest neighbors of $i$ (see Tab.~\ref{algorithms}). Hence, the neighborhoods defined in this way preserve a constant mass (\textit{i.e.,} the number of vertices is the same in all neighborhoods). Unlike for cycle and correlation networks, the adjacency matrix of the \textit{$k$-nearest neighbor network} defined in such a way is generally asymmetric, since $j\in\mathcal{N}^{(k)}_i$ does not imply $i\in\mathcal{N}^{(k)}_j$. Hence, the resulting networks are characterized by directed edges. Note that an undirected and symmetric version can easily be obtained by setting $R_{j,i}=1$ whenever $R_{i,j}=1$~\cite{Shimada2008}. 

\begin{table}[t]
\tbl{Comparison of algorithms used for constructing $k$- and adaptive nearest neighbor networks, where steps (i)-(iii) are identical for both algorithms. $S_i$ is an ordered set initially containing all the nearest neighbor indices of $i$ in increasing order of phase space distance and excluding $i$ itself, \textit{i.e.,} $\forall v < w : D_{i,S_i(v)} < D_{i,S_i(w)}$. $S_i \setminus v$ denotes the removal of nearest neighbor index $v$ from the set $S_i$, hence, $S_i(1)$ gives the index of the closest neighbor remaining in the set. For the adaptive nearest neighbor networks, a simplified algorithm is presented where the vertices are processed in temporal order, \textit{i.e.,} starting from the earliest and ending with the latest time index, as was the convention in~\cite{Xu2008}. The algorithm is readily generalized to an arbitrary processing order.}{
\centering
\begin{tabular}{p{0.4\textwidth}p{0.4\textwidth}}
\hline
$k$-nearest neighbor network & Adaptive nearest neighbor network \\
\hline
\multicolumn{2}{l}{\qquad\qquad\qquad\qquad(i) Calculate distance matrix \texttt{D(i,j)}.} \\
\multicolumn{2}{l}{\qquad\qquad\qquad\qquad(ii) Obtain \texttt{$S_i$} by implicitly sorting \texttt{i}th row of \texttt{D(i,j)}.} \\
\multicolumn{2}{l}{\qquad\qquad\qquad\qquad(iii) Initialize adjacency matrix: \texttt{A(i,j) = 0 $\mathtt{\forall}$ (i,j)}.} \\
\multicolumn{2}{l}{\qquad\qquad\qquad\qquad(iv) Fill adjacency matrix:} \\
\texttt{FOR i $\mathtt{\in \{1,\dots,N\}}$} & \texttt{FOR j $\mathtt{\in \{1,\dots,E_0\}}$} \\
\texttt{	FOR j $\mathtt{\in \{1,\dots,k\}}$} & \texttt{	FOR i $\mathtt{\in \{1,\dots,N\}}$} \\
\texttt{	}\texttt{	$\mathtt{v = S_i(1)}$} & \texttt{	}\texttt{	$\mathtt{v = S_i(1)}$} \\
\texttt{	}\texttt{	$\mathtt{A(i,v) = 1}$} & \texttt{	}\texttt{	$\mathtt{A(i,v) = A(v,i) = 1}$} \\
\texttt{		}\texttt{	$\mathtt{S_i = S_i \setminus v}$} & \texttt{	}\texttt{	$\mathtt{S_i = S_i \setminus v}$} \\ 
& \texttt{	}\texttt{	$\mathtt{S_v = S_v \setminus i}$} \\
\hline
\end{tabular}
\label{algorithms}}
\end{table}

For a $k$-nearest neighbor network, the distribution of out-degrees is always fixed at $P(k^{out})\equiv \delta(k)$. In contrast, the distribution of in-degrees allows for some variability, but must necessarily have a mean value $\left<k^{in}\right>=k$, since there are exactly $Nk$ \textit{directed} edges by definition (note that transforming the $k$-nearest neighbor network into an undirected graph~\cite{Shimada2008} leads to a network with $Nk/2$ to $Nk$ undirected edges). The remaining spatial pattern of in-degrees provides information about the attractor geometry. Specifically, we can infer that if $k_v^{in}\ll k$, $v$ lies in a phase space region with decreased density compared to the surrounding attractor. In contrast, if $k_v^{in}\gg k$, $v$ must be located in a densely populated region of the attractor.

If the coordinates of the individual vertices in the underlying phase space are known, we can consider the neighborhood size 
\begin{equation}
l_i^{max}(k)=\max_j \left\{ A_{i,j} \|\vec{x}_i-\vec{x}_j\| \right\}=\max_{j\in\mathcal{N}_i^{(k)}} \left\{ \|\vec{x}_i-\vec{x}_j\| \right\}
\end{equation}
\noindent
as a measure that is directly related with the inverse state density $\rho(\vec{x}_i)^{-1}$ of the system in the vicinity of a vertex $i$. From a statistical perspective, this strategy for retaining information about the attractor geometry can be regarded as a kernel density estimate with a simple constant kernel function, where $k$ serves as a smoothing parameter (small $k$: good spatial resolution, but large variance of the estimated state density; large $k$: small variance, but bad spatial resolution). Note that the degree centrality of an $\varepsilon$-recurrence network (see Sec.~\ref{sec:epsrecnet}) can be interpreted in a similar way, with the neighborhood size $\varepsilon$ serving as the smoothing parameter.

Figure~\ref{lorenz_linkmatrix}C displays the adjacency matrix of a $k$-nearest neighbor network for the Lorenz system, which corresponds to the respective RP (modulo the main diagonal). We also note the strong similarity with the connectivity of the associated cycle network. Reembedding the network graphs into a two-dimensional space (Fig.~\ref{lorenz_networks}C) allows recovering the double-scroll pattern of the original attractor in the reconstructed phase space of the three-dimensional embedding vectors. Note that the community structure with two ring-like network components actually reflects the different parts of the attractor.

\subsubsection{Adaptive nearest neighbor networks}

Unlike other approaches for transforming time series into complex networks, the $k$-nearest neighbor method leads to directed networks. However, in many cases the properties of undirected networks would be more directly interpretable. Moreover, the total number of undirected edges $E$ is not fixed by the algorithm itself. Specifically, there are some vertices with $k_v^{in}<k$, which has certain disadvantages if one wishes to study, \textit{e.g.,} the distributions of motifs (\textit{i.e.,} small subgraphs consisting of a fixed, low number of vertices) of a given order in the network.

In order to define an undirected nearest neighbor network with a precise control of $E$, \citet{Xu2008} as well as \citet{Small2009} suggested an alternative network construction method considering nearest neighbors but correcting for a constant number of distinct edges $E_0$ assigned to each vertex. In their approach, the network construction is an iterative process (see Tab.~\ref{algorithms}), where in each step all observations (vertices) are linked to their nearest neighbors in phase space. However, if vertex $i$ is linked with vertex $j$, vertex $i$ is removed from the neighborhood of $j$. This avoids the possibility of ``double-counting'' vertex $i$ as a neighbor of vertex $j$ and vice versa. Hence, the link between $j$ and $i$ is bi-directional, \textit{i.e.,} $A(i,j) = A(j,i) = 1$, resulting in a symmetric adjacency matrix $\mathbf{A}$, (\textit{i.e.,} an undirected network). This edge construction is repeated $E_0$ times. Finally, \textit{from} each phase space vector exactly $E_0$ edges have been drawn \textit{to} its geometric neighbors, which thus become also neighbors in a complex network sense. Consequently, there are exactly $NE_0$ undirected edges, which connect vertices of at least degree $E_0$. Specifically, a phase space vector can be a neighbor of more than $E_0$ other phase space vectors, with an average degree $\langle k \rangle = 2 E_0$. In the following, we will refer to the resulting networks as \textit{adaptive nearest neighbor networks}. 

The construction of adaptive nearest neighbor networks differs from the $k$-nearest neighbor network, since the resulting matrix is symmetric, \textit{i.e.,} the edges defined here are undirected from the beginning. Nonetheless, the process is more subtle than simply symmetrizing the recurrence matrix $\mathbf{R}$ by taking the logical matrix $(\mathbf{R}+\mathbf{R}^{T})>0$. The iterative network construction method generates a matrix such that the closest $E_0$ neighbors are always included. The exclusion process described above works to include the next closest neighbors from among the possible candidates. Note that for $E_0=k$, the adaptive neighbor network always includes all edges of the associated $k$-nearest neighbor network. However, adaptive nearest neighbor networks always have higher edge densities than $k$-nearest neighbor networks.

The frequency distribution of motifs in adaptive nearest neighbor networks has been demonstrated to be a sensitive indicator of the specific type of dynamics in the underlying dynamical system. In \cite{Xu2008} networks generated from various dynamical systems were compared and it was found that the specific distributions of the motif frequency differed qualitatively, but did so consistently. That is, periodic systems exhibit one particular type of distribution, but chaotic (one positive Lyapunov exponent) and hyper-chaotic dynamics (more than one positive Lyapunov exponent) different ones. {Specifically, motif prevalence is determined by the heterogeneity of the attractor and the intrinsic dimensionality of the system, both being larger for a chaotic system than for periodic dynamics. As a consequence, non-transitive motif patterns are more common in chaotic systems than in periodic ones~\cite{Xu2008}.} Recently, it has been shown that there is a distinct relationship between the motif distributions obtained for certain stochastic processes, and the associated scaling exponents~\cite{Liu2009b}. Note that other types of recurrence networks can be expected to show different distributions of motif frequencies. In particular, whether or not the motif distributions of other types of recurrence networks can be used for distinguishing qualitatively different dynamics as well remains a subject of future studies.

\subsubsection{$\varepsilon$-recurrence networks}\label{sec:epsrecnet}

For adaptive as well as $k$-nearest neighbor networks, choosing an equal number of neighbors for each point in phase space allows obtaining a representation of the underlying attractor that is independent of the local metric properties of the attractor in the considered embedding space. Hence, the resultant networks are based on the relative proximity between points on a trajectory in phase space and are therefore independent of any monotonic rescaling of the data (as is typically permitted by the various embedding theorems in the guise of an observational function) --- the same network will result independent of observation function. In theory this should be particularly useful for measuring metric invariants such as correlation dimension. However, it is unclear whether this method, or other types of recurrence networks are to be preferred in practice --- presumably this distinction will depend on the particular application. 

As a disadvantage of both types of nearest neighbor networks, there is no direct relationship between their local as well as global properties and the invariant density of the system under study. As an alternative, the neighborhood of a single point in phase space can also be defined by a fixed phase space distance $\varepsilon$ (see Eq.~\ref{eq_rp})~\cite{Wu2008,Gao2009,Gao2009b,Marwan2009,Donner2009}, \textit{i.e.,} by considering fixed phase space volumes instead of a fixed (local or global) number of edges. In the following, we will refer to this type of network as an \textit{$\varepsilon$-recurrence network}. Note that the structural properties of such networks often show a high degree of similarity with those of nearest-neighbor networks with similar link density (cf.~Figs.~\ref{lorenz_linkmatrix}C,D and \ref{lorenz_networks}C,D). However, the local network properties can be directly related with the phase space properties of the underlying system (for a more detailed review, see~\cite{Donner2009}). As an example, Fig.~\ref{rn_centralities_logmap} shows the trinity of centrality measures (degree, closeness, and betweenness~\cite{Freeman1979}) as well as the local clustering coefficient for realizations of the {Lorenz system} at various {values of the} control parameter {$r$}: 

\begin{figure}[thb]
\centering
\includegraphics[width=0.75\textwidth]{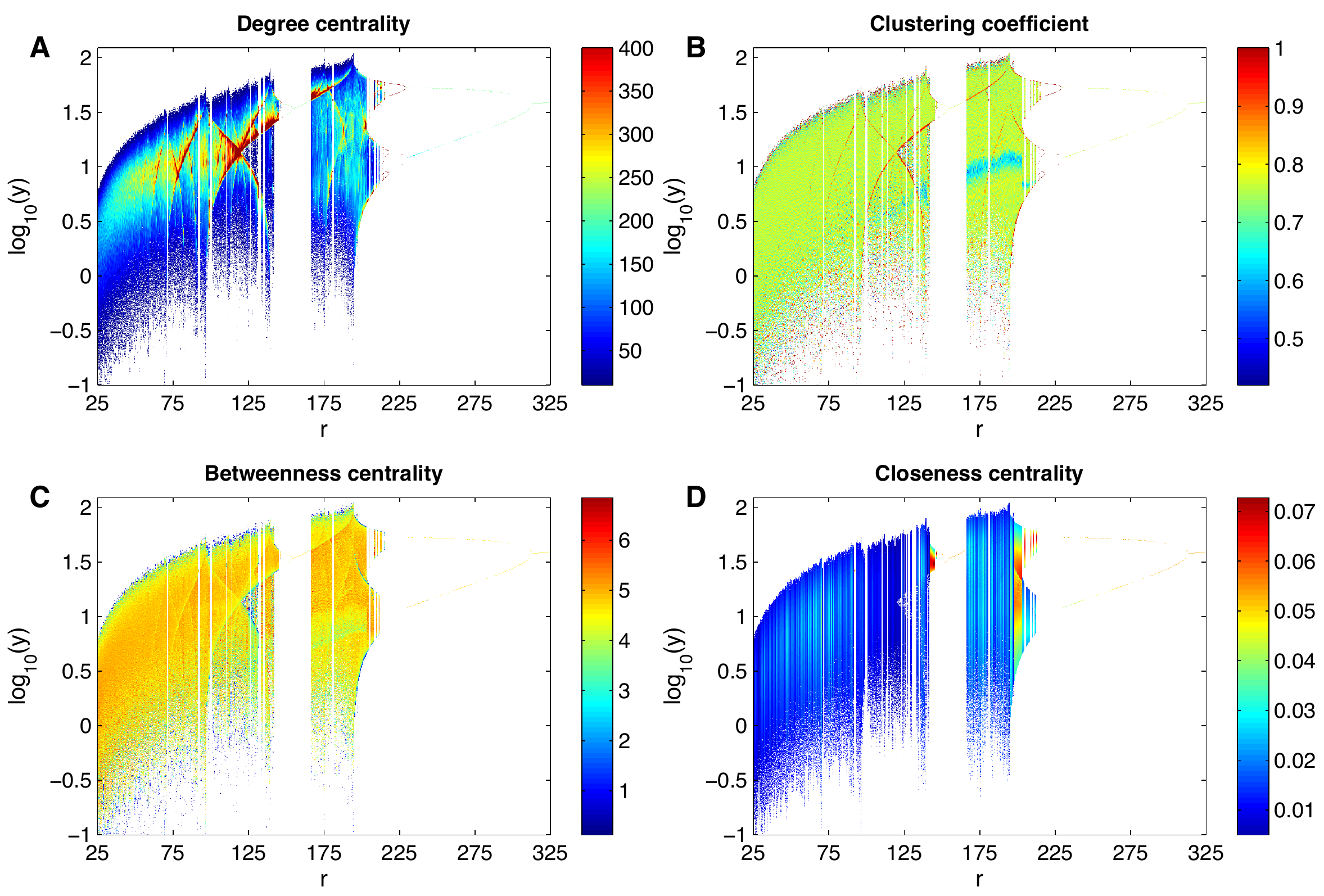}
\caption{(A) Degree, (B) local clustering coefficient, (C) betweenness (in logarithmic units), and (D) closeness centrality for the $\varepsilon$-recurrence networks obtained from trajectories of the {Lorenz system} (Eq.~(\ref{eq_lorenz})) for different control parameters {$r$. To enhance the visibility of the underlying structures, the networks have not directly been derived from the time series, but from $N=6,000$ points (for each value of $r$) in the associated Poincar\'e sections at $x=0$, $\dot{x}<0$ (maximum norm, $\varepsilon=0.05\sigma$ with $\sigma$ being the
empirical standard deviation of the considered data). $z$-coordinates have been suppressed in the figure.} Note that in contrast to the standard definition of closeness~\cite{Donner2009}, which is useful if only few isolated vertices exist, $c_v$ has been computed here separately for the individual mutually disconnected subgraphs. The broad windows in $r$ with sparse points indicate the presence of periodic orbits in the system, which have not been perfectly sampled in the Poincar\'e sections.}
\label{rn_centralities_logmap}
\end{figure}

\begin{enumerate}[(i)]

\item The \textit{degree centrality} $k_v$ (Fig.~\ref{rn_centralities_logmap}A) gives the number of neighbors of a vertex $v$ and is therefore proportional to the local recurrence rate $RR_v$ and, hence, the phase space density at the corresponding point in phase space. 

\item The \textit{closeness centrality} $c_v$ (Fig.~\ref{rn_centralities_logmap}D) is related to the inverse mean network distance of a vertex with respect to all other vertices, implying that high values of closeness can be expected in the central parts of the attractor, whereas the outer parts are characterized by small values. 

\item The \textit{betweenness centrality} $b_v$ (Fig.~\ref{rn_centralities_logmap}C) measures the number of shortest paths between pairs of vertices in the network that traverse a given vertex $v$ and, hence, indicates regions of phase space that are characterized by low density, but separate regions of higher density (geometric bottlenecks~\cite{Donner2010}). Specifically, high betweenness values indicate a strong local attractor fragmentation. 

Note that the betweenness is partially influenced be the degree: Phase space regions with high degree often show low betweenness, since there are many redundant shortest paths traversing this region. Nonetheless, betweenness centrality still yields additional information, since it is not defined exclusively locally, but encodes global network properties~\cite{Donner2009,Donner2010}. In particular, vertices with low degree, but high betweenness are of potential interest. 

We would like to remark that for a fixed $\varepsilon$, all three centrality measures are extensive network properties by definition (\textit{i.e.,} their values depend on the system size $N$, either in a linear ($k_v$) or a nonlinear ($c_v$, $b_v$) way). In contrast to this, the local recurrence rate $RR_v=k_v/(N-1)$ (\textit{i.e.,} the density of connections in the vicinity of a vertex $v$) is a non-extensive property (\textit{i.e.,} $RR_v$ does not depend on $N$ apart from possible finite-size effects). 

\item Another non-extensive vertex property is the \textit{local clustering coefficient} $\mathcal{C}_v$ (Fig.~\ref{rn_centralities_logmap}B), which measures the presence of closed triangles in the network and, hence, characterizes localized higher-order spatial correlations between observations. Specifically, since recurrence networks are spatial networks, it is possible to interpret the structures resolved by spatial variations of $\mathcal{C}_v$ in terms of the heterogeneity of the spatial filling of points. \citet{Donner2009,Donner2010} have demonstrated that this interpretation is consistent with the fact that high values of $\mathcal{C}_v$ often coincide with dynamically invariant objects, such as unstable periodic orbits or, more generally, invariant manifolds. For the {Lorenz system}, regions of the attractor with high $\mathcal{C}_v$ coincide with supertrack functions {in the Poincar\'e map (corresponding the unstable periodic orbits of the full system), \textit{i.e.,} regimes of intermittent dynamics}.

\end{enumerate}

Global network measures, such as the \textit{global clustering coefficient} $\mathcal{C}$ (\textit{i.e.,} the average value of $\mathcal{C}_v$ taken over all vertices), the closely related \textit{transitivity} $\mathcal{T}$~\cite{Boccaletti2006}, and the \textit{average path length} $\mathcal{L}$ (\textit{i.e.,} the mean graph distance between all pairs of vertices), are well suited for tracing qualitative changes in the dynamics {(see Sec.~\ref{sec:transitions})}. For the global clustering coefficient, this is a direct consequence of the different local divergence between neighboring trajectories in case of periodic and chaotic systems. However, in general, we have to carefully distinguish between discrete maps and continuous dynamical systems~\cite{Zou2010}: since the topological properties of periodic trajectories in phase space strongly differ between both types of systems, the behavior is distinctively different between maps (small $\mathcal{L}$ and large $\mathcal{C}$ and $\mathcal{T}$ for periodic orbits, large $\mathcal{L}$ and small $\mathcal{C}$ and $\mathcal{T}$ for chaotic trajectories,) and continuous systems (opposite behavior of $\mathcal{L}$ for comparable $RR$). From this perspective, we should note that (as for traditional RQA), different network measures may point to similar dynamical properties (\textit{i.e.,} may not be fully independent of each other).

{A detailed discussion of the geometric interpretation of a variety of global network properties as
well as vertex and edge properties of $\varepsilon$-recurrence networks, including graphical representations of the spatial distributions of different vertex properties for the Lorenz system in the standard parameter setting (cf. Fig.~\ref{lorenz_constr}), can be found in~\cite{Donner2009}.

\subsubsection{Recurrence network analysis and RQA}\label{sec:rnrqa}

Following the above considerations, it is evident that network-theoretic measures obtained from recurrence networks characterize (in most cases) distinctively different properties of a complex system than RQA measures. Specifically, RQA measures are based on continuous line structures in recurrence plots, \textit{i.e.,} rely on temporal interdependences between individual observations (or parts of a trajectory). In contrast to this, temporal information is not considered in the network analysis, which therefore covers mainly geometric properties of the system in phase space (\textit{i.e.,} spatial dependences). In this respect, the nonlinear statistical concept that has possibly the closest similarity with recurrence network analysis is the estimation of fractal dimensions. However, this method is much more restrictive than the network view, since it explicitly assumes the presence of geometric self-similarity in phase space. Moreover, network characteristics such as motif distributions or clustering coefficients are based on higher-order statistical dependences, \textit{i.e.,} mutual neighborhood relationships between more than two different points in phase space. In a similar way, we can argue for path-based network measures (\textit{e.g.}, average path length or betweenness centrality).

From these fundamental conceptual differences it follows that network-theoretic measures do indeed capture complementary aspects of a complex system in comparison not only to RQA, but also most other established methods of time series analysis. It should be noted, however, that there are certain fields of applications that can in principle be addressed using both traditional RQA and recurrence network analysis. One important example is the detection of dynamical transitions in time series (see Sec.~\ref{sec:transitions}). However, since both concepts provide complementary points of view, their combined use is often desirable in order to obtain additional information.
}

\subsection{Other approaches}

\subsubsection{Visibility graphs}

The concept of visibility graphs as networks of intervisible locations in physical space has been known for decades and has found many practical usages in, among other fields, engineering and urban planning~\cite{deFloriani1994,Turner2001}. Recently, Lacasa \textit{et~al.}~\cite{Lacasa2008} transferred this concept to the field of time series analysis. Here, individual observations in a time series are identified with vertices of an undirected complex network, and their connectivity is established according to a local convexity constraint between successive observations which corresponds to a visibility condition in physical space. The visibility graph approach has already found various applications~\cite{Ni2009,Lacasa2009,Liu2009,Tang2009,Yang2009,Elsner2009,Luque2009,Qian2009} and is particularly interesting for certain stochastic processes where the statistical properties of the resulting network can be directly related with the fractal properties of the time series. However, beside the relationship between the degree distribution $P(k)$ and the Hurst parameter of the underlying stochastic process~\cite{Ni2009,Lacasa2009}, convincing links between further network-theoretic measures and distinct phase space properties have not been found so far. Moreover, in its currently applied form, the visibility graph method is restricted to univariate time series analysis. {In this sense, studying the degree distribution of visibility graphs does not provide additional information, however, it may still have benefits with respect to the numerical procedures.}

As it can be seen from Fig.~\ref{lorenz_linkmatrix}E, a visibility graph typically has a distinct topology that is characterized by hubs corresponding to local maxima of the considered time series. The presence of these hubs gives rise to a pronounced community structure, where the different network clusters reflect the temporal order of observations (see Fig.~\ref{lorenz_networks}E). The mentioned general features lead to degree distributions of visibility graphs that are often found to be scale-free, which reflects the fractal properties of the underlying time series.

\subsubsection{Transition networks}

Coarse-graining the range of values in a time series into a suitable set of classes $\{S_1,\dots,S_K\}$ allows considering the transition probabilities $\pi_{\alpha,\beta}=P(\vec{x}_{i+1}\in S_{\beta}|\vec{x}_i\in S_{\alpha})$ between these classes in terms of a weighted and directed network \cite{Nicolis2005,Dellnitz2006,Gao2005,Li2006,Li2007,Gao2007,PLi2006,PLi2007,Shirazi2009,Padberg2009}. This approach is equivalent to applying a symbolic discretization with static grouping~\cite{Daw2003,Donner2008} to the phase space of the studied system. Unlike proximity networks, the resulting transition networks explicitly make use of the temporal order of observations, \textit{i.e.,} their connectivity represents causality relationships contained in the dynamics of the observed dynamical system. By introducing a cutoff $p<1$ to the transition probability $\pi_{\alpha,\beta}$ between pairs of discrete ``states'' $S_{\alpha}$ and $S_{\beta}$, we obtain an unweighted network representation, which is, however, still directed. Note that for a trajectory that does not leave a finite volume in phase space, there is only a finite number of discrete ``states'' $S_i$ with a given minimum size in phase space. This implies the presence of absorbing or recurrent states in the resulting transition network. 

The transition probability approach is well suited for identifying such ``states'' (\textit{i.e.,} regions in phase space) that have a special importance for the causal evolution of the studied system in terms of betweenness centrality $b_v$ and related measures. However, its main disadvantage is a significant loss of information on small amplitude variations. Moreover, the resulting networks do not only depend on a single parameter, but on the specific definition of the full set of classes. Note, however, that coarse-graining might be a valid approach in case of noisy real-world time series, where extraction of dynamically relevant information hidden by noise can be supported by grouping the data \cite{Daw2003}.

In contrast to the other approaches for constructing complex networks from time series, the topology of transition networks depends rather sensitively on the specific choice of discretization. For the example of the Lorenz system ($x$-coordinate) shown in Figs.~\ref{lorenz_linkmatrix}F and \ref{lorenz_networks}F, the resulting network pattern, however, reveals the spatial structure of the attractor, which is caused by the fact that for the considered relatively dense sampling of the trajectory, the transitions between subsequent observations in time always link regions of phase space that are closely neighbored. This results in the pronounced alignment of connections along the main diagonal of the adjacency matrix. However, within the two scrolls, there may also be transitions bridging several ``cells'' of the coarse-grained phase space, which is reflected by a higher connectivity of the network among the corresponding vertices. As a result, the transition network topology (represented in Fig.~\ref{lorenz_networks}F as a directed graph including self-loops) again reveals the fundamental spatial structure of the attractor.

\section{Practical considerations} \label{sec:technical}

\subsection{Cycle networks}\label{sec:technical:cycle}

For an implementation of the cycle network approach, the time series must be divided into distinct cycles. In \cite{Zhang2006,Zhang2008} the preferred method for defining cycles is splitting the trajectory at peaks (or equally troughs). In order to quantify the mutual proximity of different cycles, different measures can be applied depending on the specific application. On the one hand, the cycle correlation index $\rho_{i,j}$ (Eq.~(\ref{cyclecorr})) can be properly estimated without additional phase space reconstruction (embedding), which has advantages when analyzing noisy and non-stationary time series, \textit{e.g.,} experimental data~\cite{Zhang2006}. Moreover, this choice effectively smoothes the effect of an additive independent and identically distributed noise source~\cite{Zhang2006PRE}. On the other hand, the phase space distance $D_{i,j}$ (Eq.~(\ref{psd})) is physically more meaningful~\cite{Zhang2008}. For the example systems as well as some real-world clinical electrocardiogram recordings studied in ~\cite{Zhang2006,Zhang2008}, both methods have been found to perform reasonably well. However, whether the previously considered approaches also lead to feasible results for other cases has to be further investigated in future research.

In general, the construction and quantitative analysis of cycle networks requires a sufficiently high sampling rate, \textit{i.e.,} we require that both cycle lengths $\alpha$ and $\beta$ in Eqs.~(\ref{cyclecorr}) and (\ref{psd}) are reasonably large. The main reason for this requirement is that even two cycles that are fully identical but sampled in a different way may have rather different cycle correlation indices (and phase space distances) depending on the exact values of the observed quantity. Hence, for a very coarse sampling, it is possible that two cycles that are actually close in phase space may not be connected in the cycle network. However, for large sampling rates, the variance of this measure decreases, resulting in a more reliable network reconstruction.

\subsection{Recurrence networks}

A common problem in the construction of recurrence networks is the presence of sojourn points, which correspond to temporally subsequent observations within a small part of the phase space in the presence of strong tangential motion~\cite{gao99}. In order to avoid artificial results due to such points with strong temporal correlations, points belonging to the same ``strand'' must not be linked. As a possible solution, \citet{Xu2008} suggested that eligible neighbors should have a temporal separation greater than the mean period of the data (a considerable alternative applicable also to non-oscillatory data would be the associated correlation time). However, removing all recurrence points with a short temporal distance can lead to a loss of ``true'' recurrences as well. Moreover, we note that this approach introduces an additional parameter (the minimum recurrence time). For $\varepsilon$-recurrence networks, sojourn points can be directly removed from the complete recurrence matrix~\cite{marwan2007}. For this purpose, for every vertex $i$, those edges ($A_{i,j}=1$) with $j>i$ for which $j$ is a subsequent recurrence point of $i$ are removed ($A_{i,j}=0$). In a second step, the adjacency matrix is symmetrized again by setting $A_{j,i}=0$ whenever $A_{i,j}=0$. 

In any case, recurrence network properties depend on the sampling and possibly also the length of the time series:

\begin{enumerate}[(i)]

\item For nearest neighbor networks, using a longer time series effectively leads to a finer coverage of the available state space. As a consequence, when keeping $E_0$ (or $k$, respectively) fixed, we obtain a higher spatial resolution of the structural properties of nearest neighbor networks as the length $N$ of the time series increases. Moreover, we note that the neighborhood of a vertex will typically change with increasing $N$ since additional vertices with smaller distances in phase space appear. 

\item For $\varepsilon$-recurrence networks, the explicit choice of a threshold allows to directly control the spatial resolution. 

\end{enumerate}

In general, it is recommended to use a sampling that allows a reasonable spatial resolution of the whole phase space covered by the attractor. Moreover, since the choice of the reconstruction (embedding) parameters also matters --- as for RPs~\cite{marwan2007} --- {properly selected} embedding parameters should be used wherever possible.

\subsubsection{Adaptive nearest neighbor networks}

\begin{figure}[thb]
\centering
\includegraphics[width=0.75\textwidth]{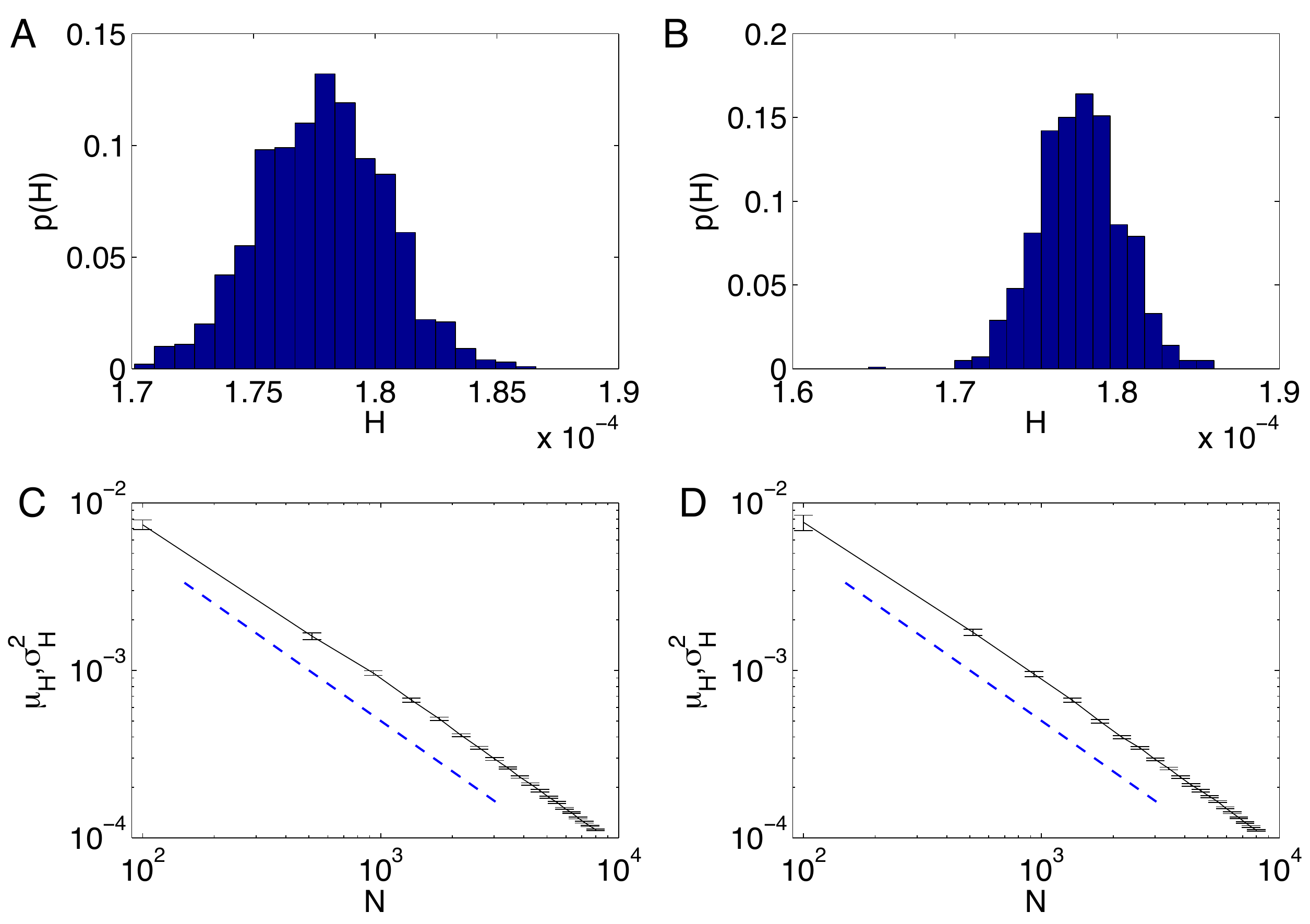}
\caption{(A,B) Probability distribution of the Hamming distance $H_{\alpha,\beta}=1-\left<\delta\left(A_{i,j}^{(\alpha)}-A_{i,j}^{(\beta)}\right)\right>_{i,j}$ (measuring the similarity between the adjacency matrices of two complex networks with the same set of vertices) between different adaptive nearest neighbor networks obtained from the same realization of (A) the R\"ossler system ($\dot{x}=-y-z$, $\dot{y}=x+ay$, $\dot{z}=b+z(x-c)$) with $a=0.15$, $b=0.2$ and $c=10$, and (B) the Lorenz system (Eq.~(\ref{eq_lorenz}), parameters as before). In both cases, $1,000$ random permutations of the same reference trajectory with $N=5,000$ data points have been used ($\Delta t=0.1$, original coordinates, supremum norm, $E_0\approx 124$ resulting in $RR\approx 0.05$). Upper bounds found obtained for vertices given in ascending order of the associated local phase space density (in comparison to the original (temporal) order of vertices) are $H^*=2.71\times10^{-4}$ and $2.53\times10^{-4}$ for the R\"ossler and Lorenz system, respectively. (C,D) Dependence of the mean value and standard deviation (error bars) of the Hamming distance (obtained from 100 permutations) on the length $N$ of the underlying time series for (C) R\"ossler and (D) Lorenz system, suggesting that $\mu_H\sim N^{-1}$ (dashed lines).}
\label{hamming}
\end{figure}

{In contrast to $k$-nearest neighbor networks, the sample network obtained from the adaptive nearest neighbor method will depend (slightly) on the order in which one processes the individual embedded time series points (originally, this was done in temporal order). Although this dependence could be completely eliminated by insisting that nodes are considered strictly in order of proximity to their neighbours there is no need to incur this additional computational complexity as the variation in the resulting network is not important. While there are many pathological situation under which small variation in the resultant networks can arise, these do not contribute to any significant structural variation in the network topology for moderate to large $N$ (see Fig.~\ref{hamming}).}

For practical applications, \citet{Xu2008} suggested that the possibility of classifying dynamical systems based on the motif distributions of adaptive nearest neighbor networks (see Sec.~\ref{sec:systemidentification}) is robust to variations in the choice of both $E_0$ (the number of edges drawn from each node) and the motif order. It is certainly true that varying $E_0$ (over a reasonable range) does not affect the corresponding results significantly. The choice of the motif order is, however, rather limited due to practical reasons --- 2- and 3-motifs offer very little scope, 5-motifs and higher orders are combinatorial nightmares and very quickly become computationally intractable. Hence, the choice of a metric based on 4-motifs is largely one of practical expedience.

\subsubsection{$\varepsilon$-recurrence networks} \label{sec:technical_rtn}

\begin{figure}[thb]
\centering
\includegraphics[width=0.6\textwidth]{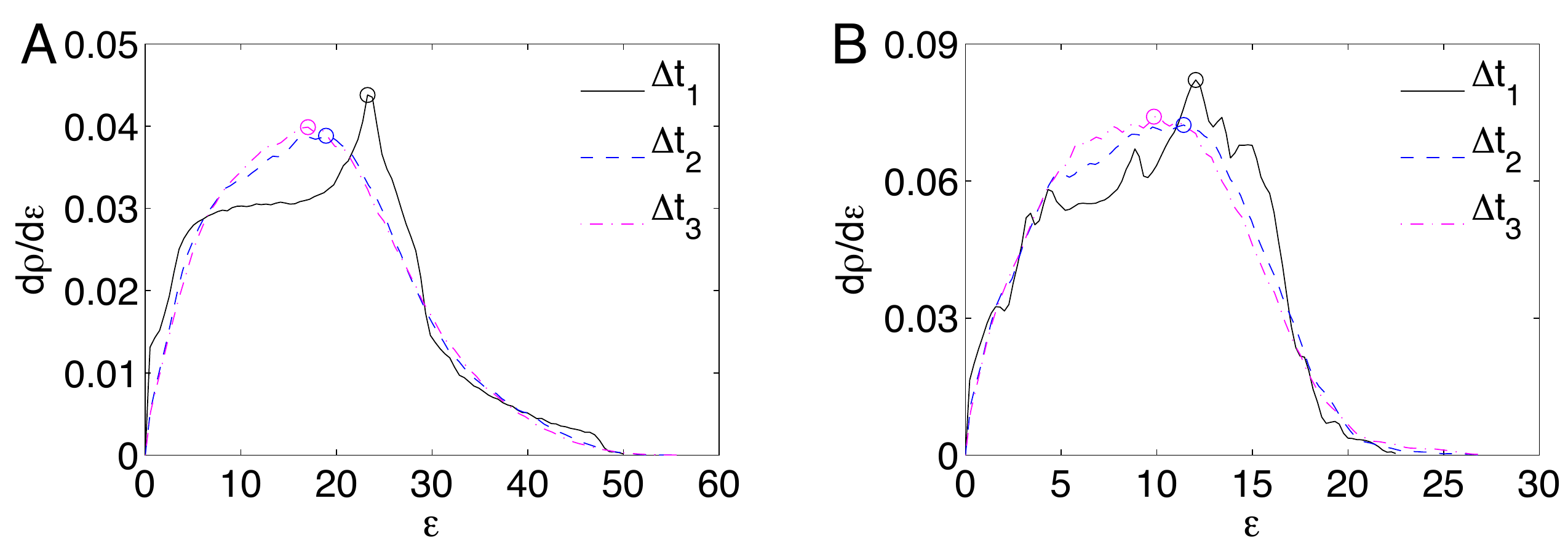}
\caption{Sampling effect on the dependence of the link density $\rho$ of an $\varepsilon$-recurrence network (equivalent to the recurrence rate $RR$) on $\varepsilon$ for (A) Lorenz system and (B) R\"ossler system with $a=0.2$, $b=0.2$ and $c=5.7$ ($N=1,000$, Euclidean norm, original coordinates). The three curves correspond to the sampling rates $\Delta t_1 = 0.01$, $\Delta t_2 = 0.1$, and $\Delta t_3 = 1.0$. Small circles indicate the recurrence thresholds suggested by the turning point criterion~\cite{Gao2009,Gao2009b}.}
\label{sampling}
\end{figure}

The problem of threshold selection has been discussed in detail by~\citet{Donner2010}, where it has been shown that simple heuristics such as the turning point criterion proposed in~\cite{Gao2009,Gao2009b} (\textit{i.e.,} determining $\varepsilon$ by the -- supposedly unique -- turning point of the $RR(\varepsilon)$ relationship) can provide misleading results. Moreover, the thresholds proposed by such general (system-independent) criteria can depend crucially on both sampling (see Fig.~\ref{sampling}) and embedding. Although there is no universal threshold selection criterion, some general considerations help fixing $\varepsilon$ at an appropriate value. On the one hand, if $\varepsilon$ is too small, there are almost no recurrence points. Hence, the information contained in the $\varepsilon$-recurrence network is rather limited. On the other hand, if $\varepsilon$ takes too large values (which is typically the case for the turning point criterion), every vertex is connected with many other vertices irrespective of their actual mutual proximity in phase space. One reasonable trade-off between these two extreme cases is choosing an $\varepsilon$ within the scaling region of the correlation integral, which coincides with the classical strategy for estimating the correlation dimension $D_2$ using the Grassberger-Procaccia algorithm~\cite{grassberger83}. Following independent arguments, \citet{schinkel2008} suggested selecting $\varepsilon$ for applications of RQA corresponding to recurrence rates $RR\leq 0.05$. Given sufficiently large $N$, this choice also provides reasonable local information about the attractor topology in phase space based on $\varepsilon$-recurrence networks~\cite{Marwan2009,Donner2009,Donner2010}.

The quantitative characteristics of $\varepsilon$-recurrence networks depend on $\varepsilon$, which is of particular importance for global network measures. Specifically, the average path length $\mathcal{L}$ is approximately inversely proportional to $\varepsilon$~\cite{Donner2009}. For other measures such as transitivity $\mathcal{T}$, global clustering coefficient $\mathcal{C}$ or assortativity $\mathcal{R}$ (\textit{i.e.,} the correlation coefficient between the degrees of all pairs of directly connected vertices), the behavior varies with the system under study. As a general observation, for large $\varepsilon$, $\mathcal{C}\to 1$ due to the increasing coverage of the attractor by the $\varepsilon$-neighborhoods. An approximate analytical theory for one-dimensional maps has been given by~\citet{Donner2009}. Corresponding statements hold for the transitivity $\mathcal{T}$ as well. The assortativity, however, shows a more diverse behavior: For $\varepsilon$ being small compared to the attractor diameter, we find a tendency towards smaller values as $\varepsilon$ increases, whereas for large $\varepsilon$, $\mathcal{R}\to 1$ since $k_v\to N-1$ for all vertices. Similar observations can be made for vertex and edge properties. However, their spatial distributions usually remain qualitatively robust as long as $\varepsilon$ does not become too large~\cite{Donner2010}. For sufficiently large $N$, the features revealed by measures, such as centralities or local clustering coefficient, can be related to finer structures in phase space for small $\varepsilon$, whereas there is a successive smoothing as the recurrence threshold increases.

When comparing different time series from the same system, it is often desirable to fix the recurrence rate $RR$ instead of $\varepsilon$. Firstly, the resulting $\varepsilon$-recurrence networks have approximately the same number of edges, which allows comparing the resulting topological properties of different networks more objectively. Secondly, the attractor diameter in phase space can change with varying control parameters. The question whether to apply a fixed $RR$ or a fixed $\varepsilon$ is especially important when cases with rather different dynamical properties are to be compared (\textit{e.g.,} periodic and chaotic orbits), where the respective $RR(\varepsilon)$ relationships are hardly comparable~\cite{Zou2010}.

\section{Applications} \label{sec:applications}

\subsection{Classification of dynamical systems} \label{sec:systemidentification}

\begin{figure}[thb]
\centering
\includegraphics[width=0.75\textwidth]{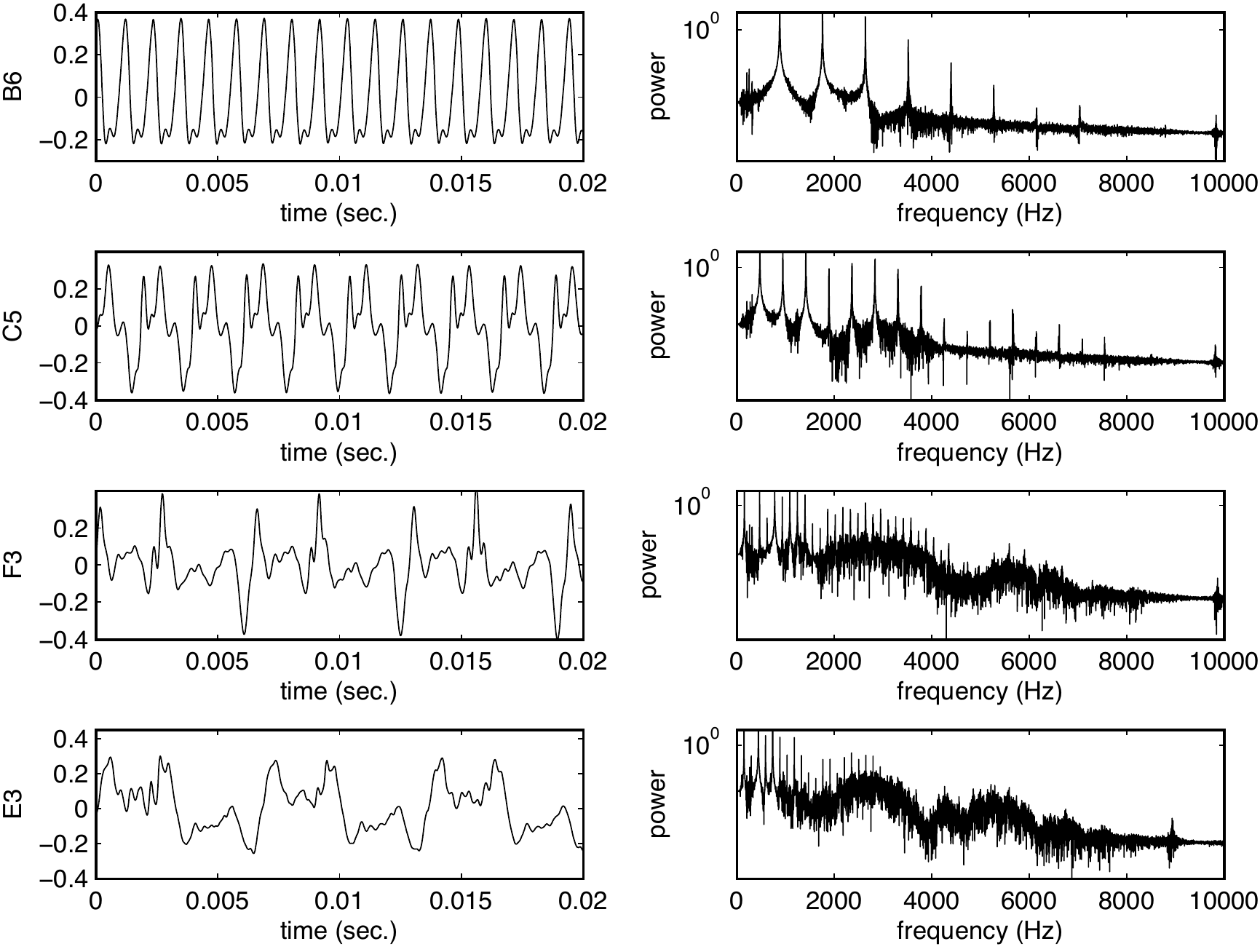} 
\caption{B$\flat$ clarinet tones, from E$_3$ (below middle C) to B$_6$ (above the treble clef), are analyzed with the adaptive nearest neighbor network method. In this figure we illustrate short sections of just three of these signals. On the left are short (20 ms) samples of the sound wave for B$_6$, C$_5$, F$_3$ and E$_3$ in the time domain. On the right we depict a corresponding sample power spectra for each. Note that the
notes E$_3$ and F$_3$ are near the bottom of the lowest (chalumeau) register of the clarinet and have a characteristic rich and woody tone. The notes C$_5$ and B$_6$ are in the intermediate clarion register (where notes are produced by overblowing --- which removes the lowest frequencies) and have a more pure, bright and penetrating sound. Notes in the altissimo register were not studied.}
\label{clardata} 
\end{figure}

\citet{Xu2008} showed that the motif prevalence in adaptive nearest neighbor networks --- in particular, the motif superfamily membership (\textit{i.e.,} the qualitative coincidence of the motif distributions within a large class of complex networks) --- can be used to classify dynamics as chaos (with one positive Lyapunov exponent), hyperchaos (multiple positive Lyapunov exponents), noise, or a periodic orbit. As a real-world example, we apply the same method to experimental data (partially depicted in Fig.~\ref{clardata}) of sustained tones voiced on a standard B$\flat$ clarinet over the dynamic range of the instrument --- from E$_3$ to B$_6$ in standard scientific notation. The 20 distinct notes where individually recorded and manually preprocessed to extract a stationary (in terms of amplitude) period of data which was then smoothed and down-sampled. Specifically, each signal was recorded at 44.1kHz and consisted of 70,000 samples from the stationary sustained phase of the intonation. This was then down-sampled to a level with approximately 25 samples per cycle. The time delay was chosen to be the first minimum of mutual information (typically between 3 and 6) and the embedding dimension was 10. From the embedded time series, adaptive nearest neighbor networks have been constructed using the Euclidean norm and four neighbors of every vertex ($E_0=4$).

\begin{figure}[thb] 
\centering
\includegraphics[width=0.5\textwidth]{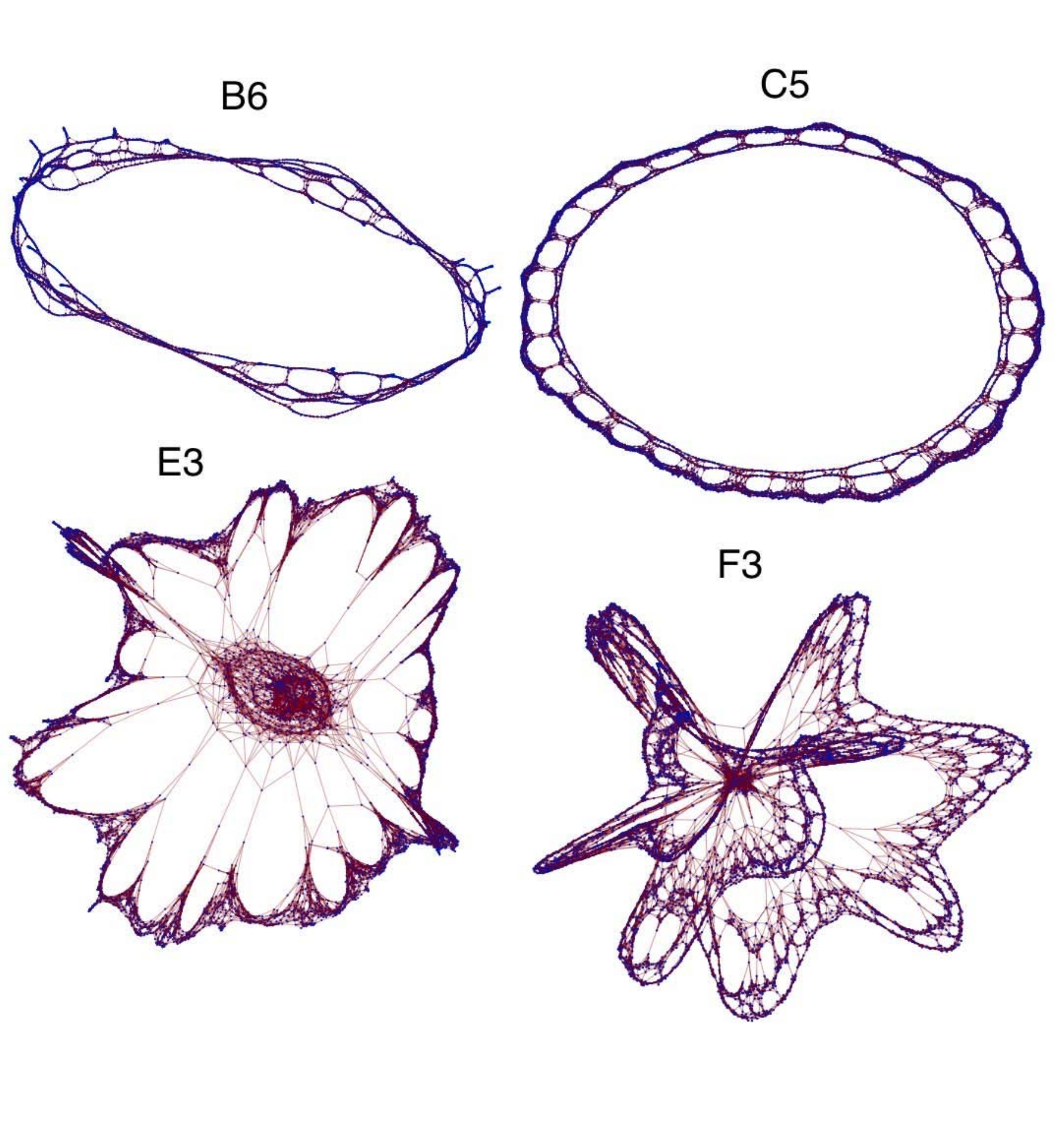}
\caption{Adaptive nearest neighbor networks for four distinct tones on the clarinet. The lower two plots E$_3$ and F$_3$ correspond to the bottom of the B$\flat$ clarinet's range (in the top half of the bass clef), the note C$_5$ is at the bottom of the clarion register (center of the treble clef) and B$_6$ is at the top.} 
\label{clarineteg} 
\end{figure}

\begin{figure}[thb] 
\centering
\includegraphics[width=0.5\textwidth]{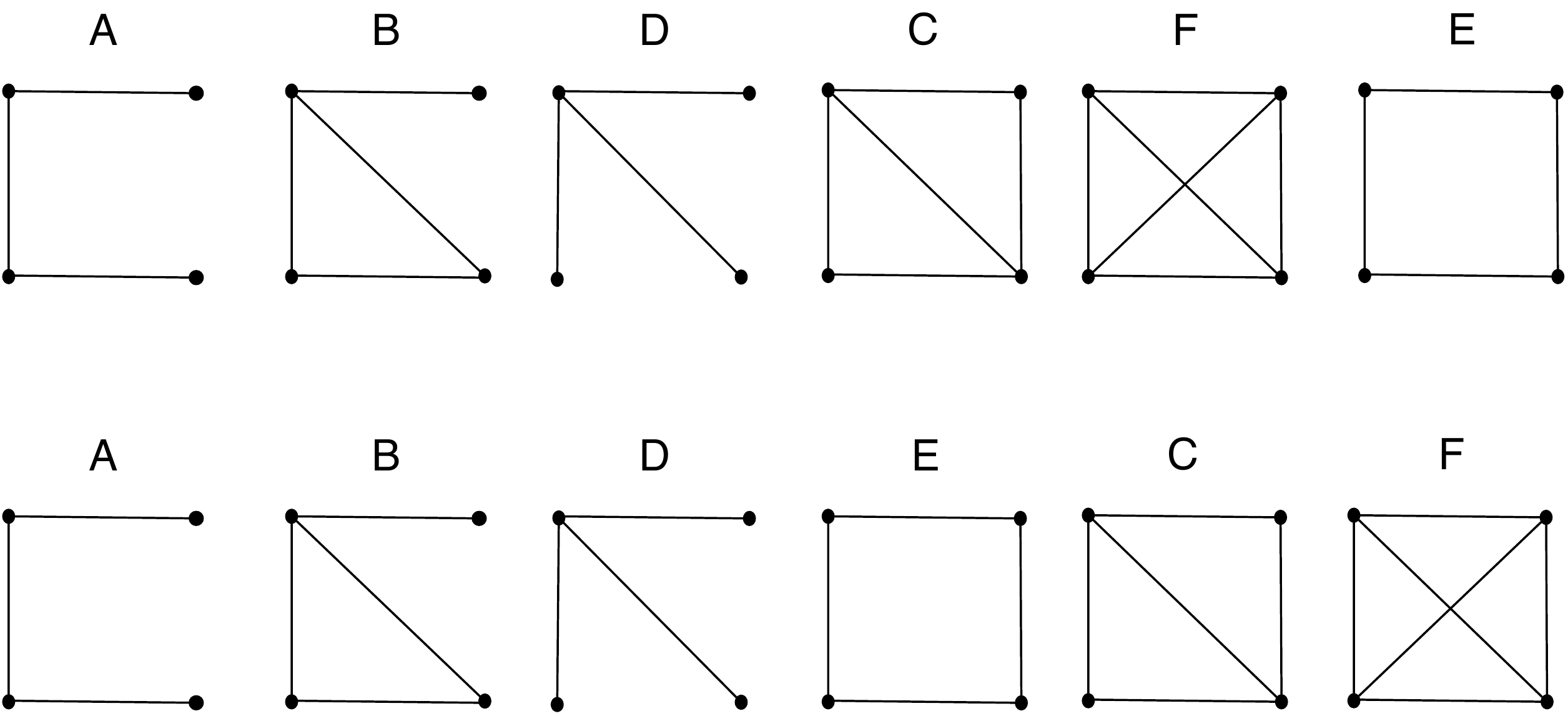}
\caption{Motif prevalence for the 20 analyzed clarinet tones (ordered according to their total frequency). Of these 20 tones, 17 belong to the same motif superfamily --- depicted in the upper row of this figure. The remaining three tones E$_3$, F$_4$ and B$_6$ belong to a distinct, but closely related motif family --- depicted in the lower row.}
\label{clarinetmotif} 
\end{figure}

Four representative networks are depicted in Fig.~\ref{clarineteg}. Despite the striking variation in the appearance of these networks (each of which is characteristic of that particular recording) we find that the mesoscopic network structures are remarkably similar. In particular, following \cite{Xu2008} we compute the frequency of motif patterns for motifs of order $4$. We find that 17 of the 20 distinct tones generate the same motif prevalence --- these tones all belong to the same motif superfamily (see top line in Fig.~\ref{clarinetmotif}). The remaining three tones belong to a distinct, but closely related family (bottom line in Fig.~\ref{clarinetmotif}).

While there is nothing concrete to link the three odd ball tones (E$_3$, F$_4$ and B$_6$), it is possible (due to the skill, or lack thereof of the musician --- or the quality of the instrument\footnote{A poor workman will blame his tools, in this case the intonation was performed by a poor musician on a cheap instrument.}) that these three tones may be prone to less stationarity. The tone E$_3$ is the lowest that the B$\flat$ clarinet will produce and the sound typically has more of a vibratory quality. The notes F$_4$ and B$_6$ are in distinct registers of the instrument, but are produced with a similar length resonance (similar fingerings) either with or without overblowing. Hence, the dynamics of these tones should be mechanistically similar. Of course, they should also be similar to many other related tones. Nonetheless, the two motif superfamilies are very similar to one another (there is only one transposition). 

The motif super-family for these three mentioned notes (E$_3$, F$_4$ and B$_6$) and the remaining seventeen notes are distinct from all those reported in \cite{Xu2008}. Nonetheless, both motif superfamilies are most similar to that of chaotic or hyperchaotic dynamics (each is only one permutation from the motif superfamily observed for hyperchaos). In all cases, the dynamics we observe in this experimental data is clearly distinct from that observed for a noisy periodic orbit.

The indication of this application --- that pure sustained clarinet tones are characteristically aperiodic, and consistent with chaotic or hyperchaotic dynamics --- is intriguing, but also preliminary. Further work with both network based methods and other techniques from nonlinear time series analysis is required.

\subsection{Identification of dynamical transitions}\label{sec:transitions}

One of the major applications of traditional RQA is the identification of dynamical transitions from time series. The RQA measures can be calculated for small square windows of size $w$ moving along the main diagonal of the RP, \textit{i.e.,} in the sub-RP $R_{i,j} |_{i,j = k}^{k+w-1}$ \cite{trulla96,marwan2007}. This approach allows studying the temporal variation of the different RQA measures, and, hence, identifying transitions in the dynamics of the studied system in terms of significant changes of these measures with time. For example, it has been shown that the diagonal line-based RQA measures are able to detect transitions between chaotic and regular dynamics in maps, whereas vertical line-based measures can identify chaos-chaos transitions~\cite{marwan2002herz}, \textit{e.g.,} in terms of detecting different properties of the laminar phases. 

Similar results have been obtained for the quantitative characteristics of $\varepsilon$-recurrence networks. {\citet{Marwan2009} studied the bifurcation cascade of the logistic map $x_{n+1}=ax_n(1-x_n)$ using both RQA measures and global properties of the recurrence networks associated with individual realizations for different values of $a$. It has been found that the presence of periodic windows is clearly detected by both transitivity (and global clustering coefficient) as well as average path length. Specifically, periodic dynamics is indicated by $\mathcal{T}=1$ ($\mathcal{C}=1$) and $\mathcal{L}=1$, whereas, for example, the recurrence rate still shows different values in dependence on the specific period. In addition, it has been observed that for finite $\varepsilon$, sudden jumps of $\mathcal{L}$ precede band merging points due to a merging of formerly disconnected network clusters. Additional pronounced minima of $\mathcal{L}$ have been found to coincide with chaos-chaos transitions. Although these bifurcations are also detectable with vertical line-based RQA measures such as laminarity or trapping time~\cite{marwan2002herz}, the shifts in the average path length are particularly well localized at the appropriate values of $a$. A similar study of complex bifurcations in a two-dimensional parameter space of the time-continuous R\"ossler system has recently been reported by~\citet{Zou2010}. A comparison with maximum Lyapunov exponents obtained from long realizations of the system revealed that recurrence network measures estimated from short time series allow a reasonable
distinction between periodic and chaotic windows, which performs somewhat better than a corresponding discrimination based on RQA measures.}

\begin{figure}
\centering \includegraphics[width=0.6\textwidth]{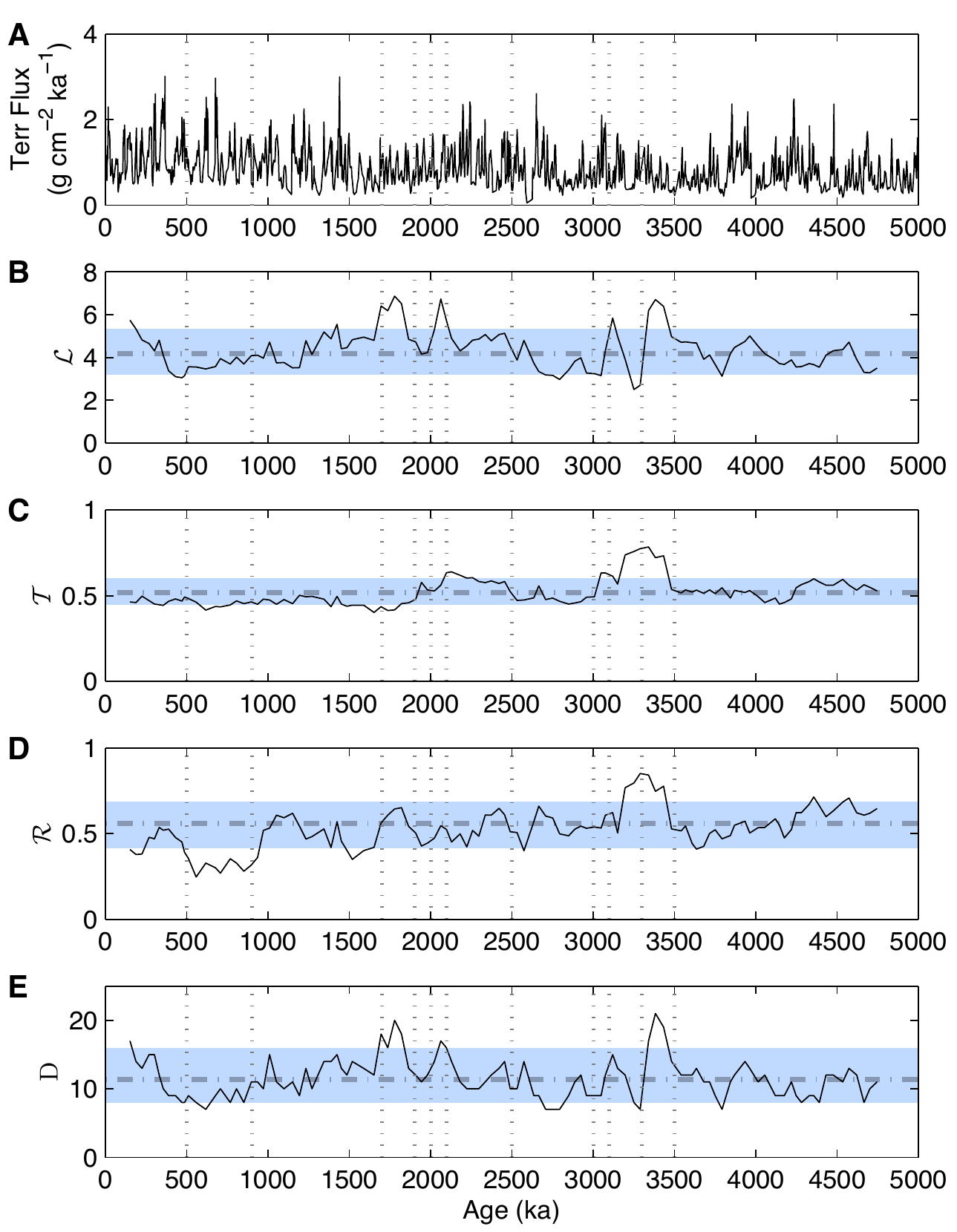}
\caption{(A) Terrigenous dust flux record of ODP site 659, and corresponding network measures (B) average path length $\mathcal{L}$, (C) transitivity $\mathcal{T}$, (D) assortativity coefficient $\mathcal{R}$ and (E) diameter $D$ obtained from $\varepsilon$-recurrence networks ($m=3$, $\tau=2$, $\rho = 0.05$, supremum norm, window size of 410~ka with 90\% overlap). The dotted vertical lines indicate {time intervals that are identified as marked features with respect to the simple confidence intervals described} in the text, the dash-dotted horizontal {lines} correspond to the mean values {for} the null-model.}
\label{fig_dust_scalar}
\end{figure}

As a real-world example for the detection of hidden transitions by means of $\varepsilon$-recurrence networks, we reconsider the analysis of a marine terrigenous dust flux record from the Ocean Drilling Program (ODP) site 659~\cite{Tiedemann1994} (see Fig.~\ref{fig_dust_scalar}A), which is located in the Atlantic close to Northwest Africa. This time series has a length of $N=1,221$, covering the last 5.0~Ma (million years) with an average sampling time of 4.1~ka (thousand years). Note that the time scale is not equidistant, the standard deviation of sampling time being 2.7~ka. Compared to the long geological time span covered, this deviation can be considered as quite small. 

The ODP 659 terrigenous dust flux record has been used to infer epochs of arid continental climate conditions and related long-term changes in the African climate. Various studies based on this record have restricted themselves to the use of linear methods of time series analysis~\cite{deMenocal1995,Ravelo2004,Trauth2009}, revealing changes in dominating cyclic (Milankovich) components and their possible relationship with known globally observable climate shifts such as the onset of Northern hemisphere glaciation, the mid-Pleistocene climate shift, or the intensification of the Walker circulation after about one million years before present (BP)~\cite{Mudelsee2005,StJohn2002,McClymont2005}.

Recently, \citet{Marwan2009} studied this time series by means of $\varepsilon$-recurrence networks. Note again that unlike most other existing methods of time series analysis, RP based techniques do not explicitly require a regular sampling. The only implicit assumption is that the data used in this approach represents the distribution of observations in the underlying phase space in a statistically reasonable way. This presumption makes RPs and recurrence networks natural candidates for the investigation of paleoclimate time series, since irregular sampling is a typical problem in the analysis of this kind of data. 

In the following, we will reconsider the recent findings and provide complementary results from additional network characteristics, using a time series that extends about 500~ka further back into the past than that used by~\citet{Marwan2009}. For consistency, we apply time-delay embedding with the optimum embedding parameters $m=3$ and $\tau=2$ ({selected based on} the false nearest neighbor and mutual information {criteria}) and a variable recurrence threshold $\varepsilon$ conserving a constant $RR=0.05$~\cite{schinkel2008,Donner2010} (cf.~Sec.~\ref{sec:technical_rtn}). To study transitions in the dust record, we construct $\varepsilon$-recurrence networks for 112 windows of size 100~samples (corresponding to periods of {on average} 410~ka) covering the line of identity ($i=j$) with a mutual overlap of 90\% (resulting in a step size of 10~samples, or approx. 41~ka). To determine the time scale of the windowed network measures, we chose the windows' mid-points.

We furthermore statistically test whether the network characteristics at a certain time differ significantly from the general network characteristics expected given the phase space distribution of state vectors for the whole embedded marine dust record and chosen window size. The corresponding null-hypothesis is that the network measures observed for a certain window are consistent with being calculated from a random draw of 100 state vectors from the prescribed phase space distribution induced by the entire time series. We can justly assume a thus randomized embedded time series without loosing essential information, because network measures are permutation-invariant (a similar test for RQA measures requires a more advanced method \cite{Schinkel2009}). In order to create an appropriate null-model, we use the following approach: 
\begin{enumerate}[(i)]
\item Randomly select {$w=100$} state vectors {$\vec{x}_{\sigma}$} from the {complete} embedded time series. {Here, the specific choice of $w$ corresponds} to the chosen window size.
\item {Use this random sample of state vectors for constructing} an $\varepsilon$-recurrence network. 
\item Calculate the network measures of interest {from this recurrence network}. 
\item Repeating this procedure $50,000$ times, we obtain a test distribution for each of the network measures. {The 5\% and 95\% quantiles of the true test distribution, which can be estimated from these empirical distributions with sufficiently high confidence, can then} be interpreted as 90\% confidence bounds (see Fig.~\ref{fig_dust_scalar}B-E).
\end{enumerate}
\noindent
{With this approach, one may test whether the spatial distribution of state vectors obtained for a given time slice is typical for the whole time series. Hence, the suggested procedure tests in fact against stationarity of certain geometric phase-space properties of the system under study. Time intervals yielding values of some network property that significantly differ from the corresponding distribution obtained from the random samples can be interpreted as possibly containing changes in the phase space structure and, hence, the observed dynamics encoded in the considered time series.}

The network measures average shortest path length $\mathcal{L}$ and transitivity $\mathcal{T}$ exhibit a distinct variability (Fig.~\ref{fig_dust_scalar}B and C). {As the most remarkable features,} $\mathcal{L}$ highlights epochs of significantly increased values between 3.5 and 3.3~Ma, around 2.1, and between 1.9 and 1.7~Ma BP\footnote{In paleoclimatology, BP (before present) refers to the number of years before the reference year 1950.}. $\mathcal{T}$ discloses epochs of increased values between 3.5 and 3.0~Ma as well as between 2.5 and 2.0~Ma. The assortativity coefficient $\mathcal{R}$ as a measure of the continuity of the density of states \cite{Donner2009} is expected to show some correlation with transitivity and global clustering coefficient. In fact, $\mathcal{R}$ also increases significantly between 3.5 and 3.2~Ma BP (Fig.~\ref{fig_dust_scalar}D). The evolution of $\mathcal{R}$ is, however, more distinct from $\mathcal{T}$ after $\sim$3.0~Ma BP, where $\mathcal{R}$ decreases markedly especially between 0.9 and 0.5~Ma BP. Moreover, there appears to be a slight trend in the evolution of $\mathcal{R}$, resulting in a tendency to decrease from higher values in the distant past towards lower values in the present. The network diameter $D$ (\textit{i.e.,} the maximum shortest graph distance between all pairs of vertices) evolves similarly to the average path length $\mathcal{L}$ (Fig.~\ref{fig_dust_scalar}E), since both measures quantify statistical properties of the distribution of shortest path lengths on the network. {Note that} the relative amplitudes of $D$ during the epochs of significant increase differ markedly from those of $\mathcal{L}$. 

The time intervals identified by the different complex network measures are robust and seem to be well correlated with some major transitions in the climate system (\textit{e.g.,} the end of the Pliocene optimum at about 3.4-3.1 Ma BP)~\cite{Marwan2009}. {We note that these specific intervals have not yet been found using classical methods of time series analysis, such as spectral analysis or breakpoint regression~\cite{Trauth2009}, and do also slightly differ from recent results obtained using RQA~\cite{Marwan2008}. We relate this to the fact that recurrence network characteristics indeed capture conceptionally different structural properties of a dynamical system from rather short time series (see Sec.~\ref{sec:rnrqa})}. A detailed climatological interpretation of our findings {will be given in an upcoming paper}.

\section{Summary}

Recurrence is a fundamental property of many dynamical processes. It is a concept successfully used in the study of dynamical and complex systems, and time series analysis, e.g., using recurrence times statistics, first return maps, recurrence plots, or recurrence quantification analysis. Conversely, network theory provides important insights in the study of many complex systems. By exploiting the duality between the recurrence matrix in the study of dynamical systems and the adjacency matrix of a complex network, we have demonstrated how information about dynamical recurrences can be used to construct complex networks from time series. These recurrence-based complex networks provide a new approach for time series analysis and offer a promising and complementary view for the study of dynamical systems. Applying well established complex network measures, we are able to characterize
and classify the dynamics of complex systems, to detect dynamical transitions, or identify invariant substructures.

{The quantitative characteristics of recurrence-based complex networks have a clear interpretation in terms of geometric properties of the underlying system in phase space~\cite{Donner2009}. In particular, recurrence networks do not take time information into account and, hence, do not explicitly rely on the presence of equally spaced observations, which is an important problem for the analysis of many real-world time series. The only implicit assumption one has to make is that the actual phase space density of the system is sufficiently well represented by the given set of points. Specifically, temporal correlations between individual observations are not taken into account, which makes recurrence-based complex networks distinctively different from the majority of other methods of time series analysis. Sufficiently long realizations guarantee stable network structures that do not depend on the specific realization of the system. However, besides the loss of information about temporal structures, the purely geometric point of view on higher-order statistical properties of a system is partially related with higher computational costs for estimating certain complex network measures (\textit{e.g.,} betweenness centrality) in comparison with other methods. Furthermore, it should be noted that there are alternative methods of complex network-based time series analysis (\textit{e.g.,} transition networks, visibility graphs, or correlation networks, cf. Sec.~\ref{sec:tsnet}), which are not based on the recurrence concept and, hence, are characterized by distinctively different properties.}

The new approach of recurrence-based complex networks combines two successful concepts in modern complex systems studies: the recurrence plot and the complex network. The first promising applications of this approach illustrate the potential of recurrence networks and their interdisciplinary relevance. {The underlying conceptual idea of constructing networks on the basis of mutual proximity relations in phase space is rather simple and thus has (as RQA~\cite{webber2009}) the potential to be applied in a meaningful way in various fields of science. We have to conclude, however, that the new method of recurrence network analysis is still in its infancy. A specific question that is necessary to be systematically addressed is which specific approach to prefer for which particular application. For example, all previous applications to well-known paradigmatic models considered only relatively low-dimensional systems, whereas recurrence network properties have not yet been explicitly studied for high-dimensional systems. We emphasize that the problem of appropriate embedding and the available amount of data can be expected to become more crucial as the dimension of the system under study increases. In summary, there are many open questions concerning the specific features and applicability of this new conceptual approach, which will surely stimulate future investigations. This statement also holds for the other recent approaches to analyzing time series by means of complex network methods.}

\section*{Acknowledgements}

This work has been financially supported by the Max Planck Society, the German Research Foundation (SFB 555 and DFG project He 2789/8-2), the Federal Ministry for Education and Research (BMBF) via the Potsdam Research Cluster for Georisk Analysis, Environmental Change and Sustainability (PROGRESS), the Leibniz association (project ECONS), a Hong Kong University Grants Council Competitive Earmarked Research Grant (PolyU 5268/07E), and a Hong Kong Polytechnic University direct allocation (G-YG35). The authors appreciate additional inspiration by fruitful comments from Peter Grassberger, Kathrin Padberg-Gehle, Jie Zhang, and Xiaoke Xu. For calculations of complex network measures, the software package \texttt{igraph}~\cite{Csardi2006} has been used. We thank K. Kramer for help with the IBM iDataPlex Cluster at the Potsdam Institute for Climate Impact Research. The network graphs shown in this paper have been created using Mathematica and the software package \texttt{GUESS} using a force directed placement algorithm (for details, see \url{https://nwb.slis.indiana.edu/community/?n=VisualizeData.ForceDirected}).

\bibliographystyle{ws-ijbc}
\bibliography{ijbcref2}

\end{document}